\documentclass{aa}

\usepackage{graphicx}
\usepackage{txfonts}
\usepackage{hyperref}
\usepackage{gensymb}

\begin{document}

   \title{Discovery and analysis of low-surface-brightness galaxies in the environment of NGC~1052}

   \titlerunning
   {LSBGs in the NGC~1052  environment}
   \authorrunning{Román et al.}

   \author{Javier Rom\'an\inst{1,2,3}\thanks{   \email{jromanastro@gmail.com}}
   \and
   Aida Castilla\inst{4}
   \and
   Javier Pascual-Granado\inst{1}
   }

   \institute{Instituto de Astrof\'isica de Andaluc\'ia (CSIC), Glorieta de la Astronom\'ia, 18008 Granada, Spain
   \and Instituto de Astrof\'{\i}sica de Canarias, c/ V\'{\i}a L\'actea s/n, E-38205, La Laguna, Tenerife, Spain
   \and Departamento de Astrof\'{\i}sica, Universidad de La Laguna, E-38206, La Laguna, Tenerife, Spain
   \and Universidad Internacional de Valencia (VIU), Pintor Sorolla 21, 46002 Valencia (Spain)
   }
   \date{\today}

 
  \abstract
   {The environment of NGC~1052 has recently {attracted} much attention because of the presence of low-surface-brightness galaxies (LSBGs) with apparently ``exotic'' properties, making it a region of high interest for the detection of new objects. We used public deep photometric data from the Dark Energy Camera Legacy Survey to carry out a comprehensive search for LSBGs over a wide region of 6$\times$6 degrees, equivalent to 2$\times$2 Mpc at the distance of NGC~1052. We detected 42 LSBGs with r$_{eff}$~>~5~arcsec and $\mu_{g}(0)$~>~24~mag~arcsec$^{-2}$, of which 20 are previously undetected objects. Among all the {newly} detected objects, RCP~32 stands out with extreme properties: r$_{eff}$~=~23.0~arcsec and <$\mu_{g}$>$_{eff}$~$=$~28.6 mag arcsec$^{-2}$. This makes RCP~32 one of the {lowest} surface brightness galaxies ever detected through integrated photometry, located at just 10 arcmin from the extensively studied NGC~1052-DF2. We explored the presence of globular clusters (GCs) in the LSBGs. We marginally detected a GC system in RCP~32, and argue that this {LSBG} is of great interest for follow-up observations given its extremely low baryon density. After analyzing the distribution of galaxies with available spectroscopy, we identified a large-scale structure of approximately 1~Mpc that is well isolated in redshift space and {centered on} NGC~1052. The spatial correlation {analysis} between the LSBGs and this large-scale structure suggests their association. However, when exploring the distribution of effective radius, we find an overpopulation of large LSBGs (r$_{eff}$~>~15~arcsec) located {close to} the line of sight of NGC~1052. We argue that this is suggestive of a substructure with similar radial velocity in sight projection, but at a closer distance, to which some of these apparently larger LSBGs could be associated. {However. possible effects derived from tidal interactions are worthy of further study}. Our work expands the catalog of LSBGs with new interesting objects and provides a detailed environmental context for the study of LSBGs in this region.
}



   \maketitle
%

\section{Introduction}

Low-surface-brightness galaxies (LSBGs) are the dominant population of the faint end of the galaxy luminosity function \citep[e.g.,][]{2005ApJ...631..208B, 2012AJ....143..102G, 2019MNRAS.485..796M}. Due to their low baryon density, LSBGs are an excellent laboratory for testing galactic formation and evolution models \citep[][]{1987AJ.....94...23B, 1989ApJ...344..685K, 2001MNRAS.321..559B, 2001ApJ...552L..23D, 2020MNRAS.494.1848S} and provide important observational constraints {for} the $\Lambda$-CDM cosmological paradigm \citep[e.g.,][]{1999ApJ...524L..19M, 1999ApJ...522...82K, 2017ARA&A..55..343B}. Traditionally, defined as those galaxies fainter than the surface brightness of the night sky, $\mu_V>$~22~mag~arcsec$^{-2}$ \citep[][]{1984AJ.....89..919S, 1988ApJ...330..634I, 1988AJ.....96.1520F, 1991ApJ...376..404B, 1997AJ....114..635D}, LSBGs are currently
 broadly defined as galaxies fainter than the surface brightness detection limit of the Sloan Digital Sky Survey for objects of  small {apparent} size, that is approximately 24~mag~arcsec$^{-2}$ in the \textit{g} band. 

{Because of their inherent low surface brightness, the detection of new LSBGs is severely limited by depth and data processing, where correct flat-fielding and sky subtraction are crucial steps.} Continuous improvement in the design and construction of deep multi-purpose optical surveys such as the IAC Stripe82 Legacy Survey \citep[][]{2009ApJS..182..543A, 2016MNRAS.456.1359F, 2018RNAAS...2..144R}, the Dark Energy Camera Legacy Survey \citep[DECaLS;][]{2016MNRAS.460.1270D}, the Hyper Suprime-Cam Subaru Strategic Program \citep[HSC-SSP][]{2018PASJ...70S...4A} and more is bolstering active research in the field of LSBGs \citep[e.g.,][]{2015ApJ...798L..45V, 2015ApJ...807L...2K, 2017MNRAS.468.4039R, 2017A&A...608A.142V, 2018ApJ...857..104G, 2019MNRAS.488.2143P, 2021ApJS..252...18T} and the number of new detected LSBGs is increasing dramatically. However, the possibility of accessing data with lower surface brightness limits is not enough to obtain an unbiased view of this low-mass and low-brightness galactic population, and their study still presents numerous challenges. For instance, traditional computational utilities for automated source detection and analysis are usually not efficient for low-surface-brightness objects. This implies the need for specialized software to perform detection and photometry that maximizes the potential provided by the data \citep[see a review by][]{2021A&A...645A.107H}. Another aspect is related to automatic morphological classification and the identification of artifacts that parametrically mimic LSBGs, such as reflections, faint clumped sources, and inaccurate deblending identifications. Human visual inspection allows the vast majority of these false positives to be removed with high efficiency, but for large data sets this is a very tedious and even unaffordable task. While the use of deep learning procedures has recently proven efficient in the identification of false positives \citep[e.g.,][]{2018MNRAS.475..894T, 2019MNRAS.490.3952B, 2021A&C....3500469T} with {an accuracy} of around 90\%, the {accuracy} of human visual identification {has not yet been reached}. 

Obtaining radial velocities through optical spectroscopy for complete samples of LSBGs is prohibitive in terms of observational time, and objects not resolved into stars (thus outside the scope of star-counting techniques), those lacking HI gas, and those with extremely low surface brightness ($\mu$ > 26 mag arcsec$^{-2}$) are particularly problematic. This has restored the importance given to traditionally secondary methods in estimating distances, such as the surface brightness fluctuation method \citep[e.g.,][]{2019ApJ...879...13C, 2021ApJ...908...24G, 2021MNRAS.504.1668Z} and to the use of the peak of the globular cluster (GC) luminosity function as a standard candle \citep[e.g.,][]{2019MNRAS.486..823R}. 

With the increasing depth of available photometric data, and therefore the diminishing surface brightness of the detected objects, the study of LSBGs is becoming more and more challenging. {A need exists} to develop increasingly advanced analysis techniques in the face of the imminent arrival of the new generation of deep optical and infrared surveys such as the Legacy Survey of Space and Time \citep[LSST; ][]{2009arXiv0912.0201L}, Euclid \citep[][]{2011arXiv1110.3193L} or the Nancy Grace Roman Space Telescope \citep[][]{2015arXiv150303757S}, among many others.

In this work we aim to explore the presence and properties of LSBGs in the environment of the NGC~1052 group of galaxies. This region is of particular interest in light of the ``exotic'' properties that recent works claim {for} some LSBGs found in this region. {For instance,} \cite{2018Natur.555..629V} in [KKS2000]~04 (more commonly known as NGC~1052-DF2) and \cite{2019ApJ...874L...5V} in NGC~1052-DF4 claimed the similarity between the observed baryonic matter and the dynamic matter obtained by the radial velocity dispersion of their GCs and also the stellar component \citep[][]{2019A&A...625A..76E, 2019ApJ...874L..12D}. {This} observational evidence would imply an extreme and unexpected deficit (even lack) of dark matter in these LSBGs. Taking into account that their metallicity content follows the usual stellar mass--metallicity relation for dwarf galaxies \citep[][]{2019A&A...625A..77F, 2019MNRAS.486.5670R}, the possibility that they are tidal dwarf galaxies -- galaxies formed by strong interactions with recycled material from massive host galaxies, high in metals, and intrinsically born with a lack of dark matter \citep[see e.g.,][]{2012ASSP...28..305D, 2021A&A...649L..14R} -- is ruled out. The properties of these LSBGs have been the subject of much debate \citep[see e.g.,][]{2018ApJ...859L...5M, 2018MNRAS.480L.106O, 2018MNRAS.480..473F, 2018Natur.561E...4K, 2020MNRAS.491L...1L, 2019MNRAS.489.2634H, 2019A&A...624L...6M, 2019MNRAS.484..510N, 2020ApJ...904..114M, 2021ApJ...919...56M}. One hypothesis put forward by \cite{2019MNRAS.486.1192T} is that these LSBGs "lacking dark matter" could be at a closer distance than initially estimated by \cite{2018Natur.555..629V, 2019ApJ...874L...5V}. This would not only make the stellar mass smaller (leaving room for a higher M/L fraction and therefore recovering a certain amount of dark matter, alleviating its strong deficit) but would also resolve the other anomalies that these LSBGs have related to their globular cluster luminosity function {(GCLF)} \citep[e.g.,][]{2021ApJ...909..179S}, something considered a universal property of any galactic population, including LSBGs \citep[e.g.,][]{2007ApJS..171..101J, 2010ApJ...717..603V, 2012Ap&SS.341..195R, 2018MNRAS.475.4235A, 2019MNRAS.484.4865P}. However, the distance to these galaxies is still the subject of intense debate \citep[][]{2018ApJ...864L..18V, 2018RNAAS...2..146B, 2019ApJ...880L..11M, 2019RNAAS...3...29V, 2020ApJ...895L...4D, 2021ApJ...914L..12S, 2021MNRAS.504.1668Z} and there is currently no broad consensus on these distance values, with the {remaining debated range of distances being approximately 12--22 Mpc}.

{Therefore, the motivation for the present study is twofold. On the one hand, we perform a systematic detection of LSBGs with the best photometric data available in the environment of the NGC~1052 galaxy group, aiming to find new potentially interesting objects given the presence in this region of LSBGs with exotic properties. On the other hand, we carry out an exploration of the structures that exist in this region in order to delimit the environment and study its properties together with the sample of LSBGs.}

This article is structured as follows: in Section 2 we describe the data used, in Section 3 we present the sample and its properties. In Section 4 we carry out the analysis of the sample and in Section 5 we discuss the results. In this work we use the AB photometric system. {All photometric magnitudes are extinction corrected \citep{2011ApJ...737..103S}.} We use standard cosmology with $\Omega_m$~=~0.3, $\Omega_\Lambda$~=~0.7, and H$_0$~=~70 km s$^{-1}$ Mpc$^{-1}$. The graphics of the figures are optimized for viewing in the electronic version.

\section{Data}\label{sec:data}

We used photometric data from the Dark Energy Camera Legacy Survey \citep[DECaLS;][]{2016MNRAS.460.1270D}. This survey makes use of the Blanco 4m telescope located at the Cerro Tololo Inter-American Observatory, which provides {wide-field} multiband photometry in \textit{g}, \textit{r,} and \textit{z} bands with a median seeing of 1.5 arcsec {using the Dark Energy Camera \citep[DECam;][]{2008arXiv0810.3600H}. Data obtained with this instrumentation have proven to be very useful for searching for LSBGs \citep[e.g.,][]{2017A&A...597A...7M, 2021ApJS..252...18T, 2021A&A...652A..48M}.} Detailed information about DECaLS and related surveys is provided by \cite{2019AJ....157..168D} and on the Legacy Survey website\footnote{https://www.legacysurvey.org/}. We used publicly available data from the seventh data release (DR7).

The depth of the data in the environment of NGC~1052 is remarkable, however we found that it is not homogeneous, varying from 29.2 to 29.9 mag arcsec$^{-2}$ in the \textit{g} band, from 28.7 to 29.1 mag arcsec$^{-2}$ in the \textit{r} band, and from 27.5 to 27.6 mag arcsec$^{-2}$ in the \textit{z band}, all measured as 3$\sigma$ in 10$\times$10 arcsec boxes following the depth definition by \cite{2020A&A...644A..42R}, Appendix A. This range in depth is from approximately 0.5 to 1 mag arcsec$^{-2}$ deeper than the typical depth of the DECaLS data in other regions \citep[see][]{2021arXiv210406071M}. It is also considerably deeper than other existing data sets in this region, for example the Sloan Digital Sky Survey (27.4 and 26.8 mag arcsec$^{-2}$ in the \textit{g} and \textit{r} bands respectively) or the Dragonfly Telephoto Array \citep[28.9 and 28.2 mag arcsec$^{-2}$ in the \textit{g} and \textit{r} bands, respectively;][]{2014PASP..126...55A, 2016ApJ...833..168M} all measured as 3$\sigma$ in 10$\times$10 arcsec boxes. This makes the DECaLS data the most optimal for this work, approximately 1 mag deeper in g and r bands than previous works\footnote{{But see very recent work by \cite{2021arXiv210909778K} in the environment of NGC~1052 and \cite{2021A&A...654A..40T} in the NGC~1042 galaxy.}}, covering a much larger area, with better seeing and with the additional z band.

In order to explore a region wide enough to cover the environment around NGC~1052 we selected an analysis area of 6$\times$6 degrees, corresponding to 2.07$\times$2.07 Mpc at the distance of NGC~1052, which is of 20 Mpc \citep[][]{2013AJ....146...86T}. {We thus} explore an environment of at least 1 Mpc around NGC~1052 at its distance \citep[the virial radius of NGC~1052 has been calculated to 390 kpc by][]{2019MNRAS.489.3665F}. In coordinates this region comprises 37.37º~<~R.A.~<~43.27º and -11.26º~<~Dec.~<~-5.26º. We created these 6$\times$6 degree mosaics for each of the \textit{g}, \textit{r,} and \textit{z} bands by merging individual ``co-add'' fields downloaded by querying the survey archive using a dedicated pipeline.


\section{The sample}

\subsection{Detection of low-surface-brightness galaxies}\label{sec:detection}

{The detection of LSBGs is usually carried out by a sequence of tasks}. The first step is a broad detection of LSBG candidates. Both the depth of the data and the efficiency of this detection {in} the low-surface-brightness regime will define the completeness of the sample, and so a high efficiency in the detection of diffuse sources in this first step is a key point in the process. Given the importance of this primary detection, it is common {to} use specialized software or procedures \citep[e.g.,][]{2015ApJS..220....1A, 2018MNRAS.478..667P, 2021A&A...645A.107H}. After a first detection of potential objects, a characterization of their structural and morphological properties is necessary in order to accurately define the sample, typically with certain criteria in surface brightness and radius. To obtain the structural parameters of the LSBG candidates, the detected sources are usually fitted with a S\'ersic model \citep{1968adga.book.....S}. Specific procedures such as \texttt{GALFIT} \citep{2010AJ....139.2097P} or \texttt{IMFIT} \citep{2015ApJ...799..226E} are {used to obtain accurate} morphological and structural parameters. However, the computational cost of these is very high, becoming a bottleneck in LSBG analysis pipelines. For this reason, it is advisable to minimize the number of LSBG candidates prior to their S\'ersic fitting to save computational time. This is not trivial because structural parameters from automated segmentation catalogs are much less accurate than S\'ersic fit modeling, making correct screening of the sample challenging. As a last step, {a ``supervision''} of all the objects matching the criteria of the sample is necessary in order to discard the frequent presence of false positives. These are sources that, although meeting the imposed criteria, are clearly not LSBGs. Examples of false positives are clumped sources that have been considered as a single source, faint reflections due to the optics of the instrumentation, or artifacts present in the images. Depending on the data volume analyzed, the number of objects to be supervised can become very high. For this reason, it is increasingly common to use deep learning techniques to screen objects in order to eliminate these false-positive detections \citep[e.g.,][]{2018MNRAS.475..894T, 2019MNRAS.490.3952B, 2021A&C....3500469T}. However, unsupervised or automated detections can not yet reach the {accuracy} that human visual inspection is capable of obtaining, reaching almost 100\%. Therefore, human supervision is desirable, when possible, in order to {minimize the presence of false positives in}  samples of reasonably limited volume, such as that used here.

\begin{figure}
        \includegraphics[width=1.0\columnwidth]{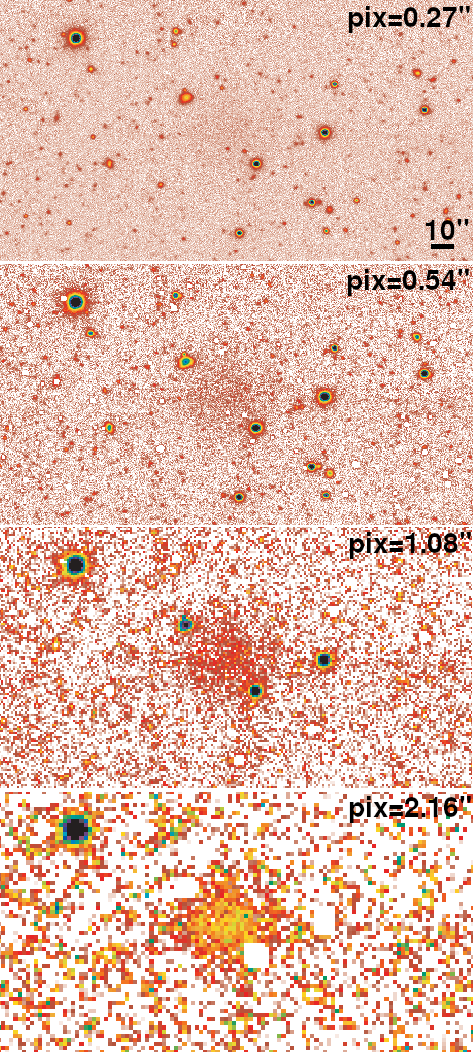}
    \caption{Example of sequential masking and rebinning centered in the lowest surface brightness galaxy of the sample (<$\mu_{g}$>$_{eff}$~$=$~28.6~mag~arcsec$^{-2}$). Each panel shows a stage of the rebinning process. The image detection limit is 29.9 mag arcsec$^{-2}$ at 3$\sigma$ measured in 10$\times$10 arcsec boxes (\textit{g} band).}
    \label{fig:Binning}
\end{figure}

Taking these considerations into account, we designed a pipeline for the detection of LSBGs prior to  S\'ersic
modeling, the most computationally expensive process. The aim of this pipeline is twofold. On the one hand, it performs a quality filtering of all sources, selecting only those compatible with LSBGs, that is, extended sources with low surface brightness. On the other hand, this procedure efficiently detects objects of extremely low surface brightness. The operation of this procedure is based on the reduction of Poissonian noise by rebinning the images after masking high-surface-brightness objects. It allows the signal-to-noise ratio (S/N) to be increased, which in turn increases the detectability, by \texttt{SExtractor,}  of diffuse sources buried in the Poissonian noise \citep{1996A&AS..117..393B} after a significant loss of resolution. Given the large radius expected for the LSBGs in our sample, there is a wide margin to carry out this loss-of-resolution process by rebinning the pixel scale. This method of enhancing the diffuse emission by averaging small spatial scales has been used in previous works with {good} results \citep{2019ApJS..240....1Z, 2020A&A...644A..42R}. The operation of this pipeline is the following: We create a deep \textit{g} + \textit{r} image that is going to be used as a detection image. In this image, we run \texttt{SExtractor} with a set of parameters designed for the detection of small size and high surface brightness sources and we mask all these sources in the image. Next, we rebin the masked images to 2$\times$2 pixels, in which each pixel in the rebinned image is the average of the four original pixels, not taking into account masked values. This sequence of detecting, masking, and rebinning is repeated three times. Starting from the original image of 0.27 ''/pixel, it produces a final 2.16~''/pixel image in which the diffuse emission is enhanced with a significant gain of S/N (factor 8) with a consequent loss of resolution. We note that objects with low surface brightness do not run the risk of being masked in this process, because the detection threshold is defined well above the typical surface brightness of these objects. In Fig. \ref{fig:Binning} we show the results of this computational procedure for the different sequences or pixel scales, centered on the faintest LSBG detected in our work (<$\mu_{g}$>$_{eff}$~$=$~28.6~mag~arcsec$^{-2}$).

We then ran \texttt{SExtractor} in dual mode {on} these enhanced images. In order to detect potential sources with a small radius, we use both 2.16~''/pixel and 1.08''/pixel images {for} the detection.  We used
the \textit{g} + \textit{r} mosaic  as detection images in order to measure the parameters in
the \textit{g} and \textit{r} bands separately. At this stage, we discard the use of the much shallower \textit{z} band. Among different parameters, we obtained the maximum surface brightness of the sources present in these enhanced images. These parameters, being obtained with \texttt{SExtractor}, do not have the precision that a S\'ersic fitting provides, but given the good masking of the aforementioned process, the obtained values are robust enough to perform a source screening.  To select LSBG candidates, we selected sources with a conservative value of $\mu_{g,max}$ = 23 mag arcsec$^{-2}$ according {to} the value provided by \texttt{SExtractor}. This criterion provides a margin given that the final criterion to select the LSBGs is a central surface brightness fainter than 24 mag arcsec$^{-2}$ in the \textit{g} band, obtained from the more accurate S\'ersic fitting. The number of selected LSBG candidates matching this criterion is 335 sources. It is worth noting the great computational efficiency of this procedure: the total computational time for this detection in the 6$\times$6 degrees mosaic is approximately 1 hour using an AMD Ryzen 5 3600 six-core processor. 

Additionally, we carried out {an independent} visual detection of LSBGs in the images. For that we used the great image quality in color composed images provided by the Legacy Survey browser\footnote{\url{https://www.legacysurvey.org/viewer}} together with the enhanced images previously discussed. The detection was carried out {by one member of the team} by selecting sources with low surface brightness but conservatively, this is selecting sources that could definitely have a higher surface brightness than the final criteria to be applied, so as not to introduce any bias that could influence detection. As the aim of this detection is to find candidates that will finally be fitted to obtain their structural parameters with great precision, we do not make any cut in radius, that is, we exclusively select sources with diffuse brightness, again so as not to introduce any bias in this regard. The total number of sources acquired by visual inspection was 2268, {leading to} a total number of 2603 LSBGs candidates.

We then obtained accurate structural and photometric parameters by fitting a S\'ersic model to all 2603 candidates. For this we use the software \texttt{IMFIT} in stamps centered on the candidate sources in which external sources have previously been masked. These stamps have the original data resolution of 0.27~''/pixel and the masking is performed aggressively in order to hide diffuse light from external sources. The fitting is performed in each of the available \textit{g}, \textit{r,} and \textit{z} bands using a single S\'ersic model with point spread funcion (PSF) deconvolution and variable sky background. The position angle and ellipticity of the target are calculated in the deep \textit{g} + \textit{r} image beforehand, {and are} then introduced as fixed parameters in the fitting of the individual \textit{g}, \textit{r,} and \textit{z} bands. To obtain the PSF model, we used a similar procedure to that by \cite{2020A&A...644A..42R} and \cite{2020MNRAS.491.5317I}. This consists of making use of several unsaturated stars that after normalization and combination produce a PSF model. As the objects to be PSF deconvolved (LSBG candidates) will be relatively small in size, we created a PSF model of 10 arcsec in radius. This is more than enough to correct for the effect of PSF on these extremely low-surface-brightness objects. 

\begin{figure*}
\centering
        \includegraphics[keepaspectratio,width=0.97\textwidth]{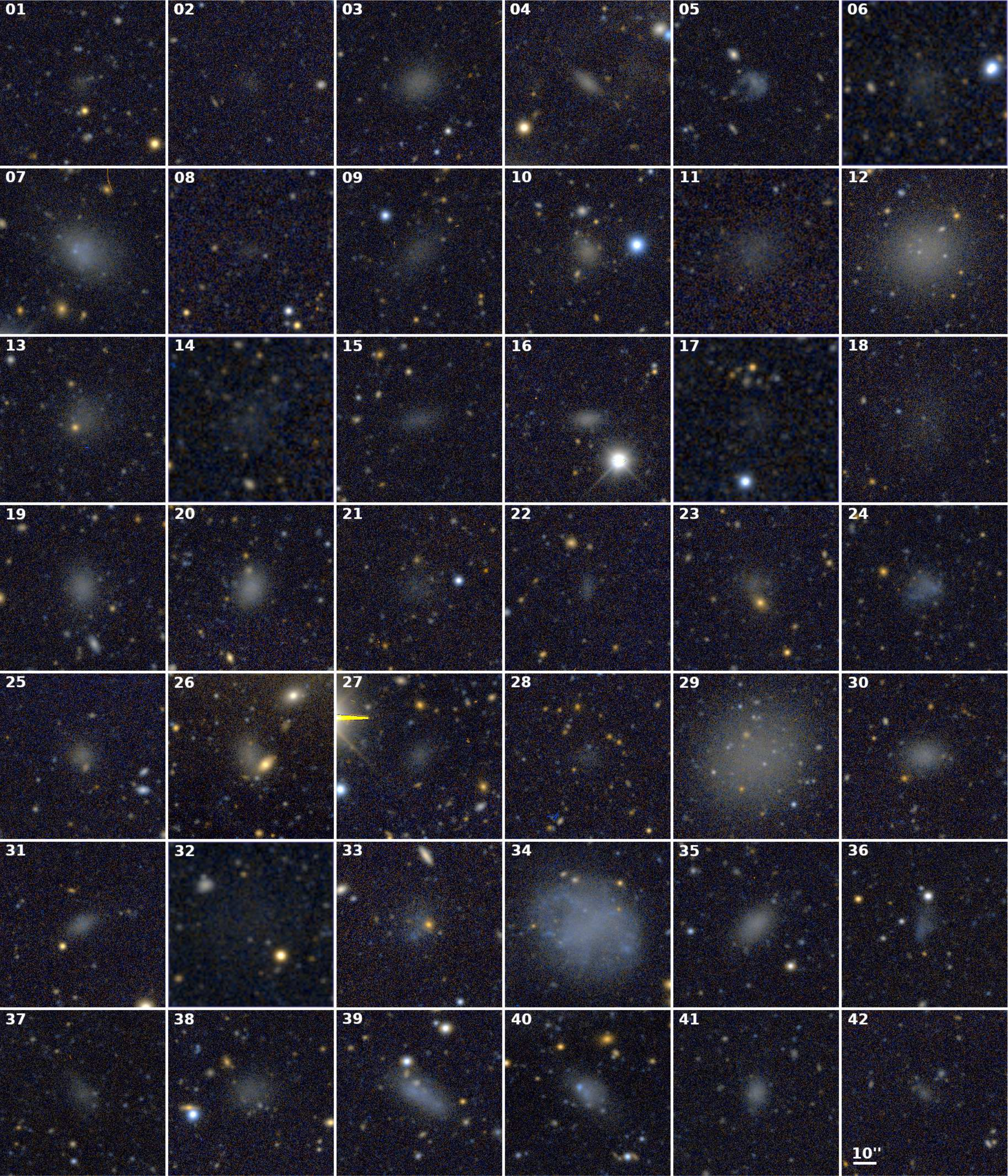}
    \caption{Color-composed images with \textit{g} and \textit{r} bands of all LSBGs in the sample. The size of the stamps is 81 arcseconds, and north is up and east is to the left. The images of the fainter galaxies have been smoothed with a Gaussian kernel to enhance contrast. {The ID catalogue number (see Table \ref{tab:IDs}) is placed in the upper left corner of each LSBG.}}
    \label{fig:Ap1}
\end{figure*}

The selection criteria for the final sample of LSBGs are a central surface brightness of $\mu_g(0)$~$>~$~24~mag~arcsec$^{-2}$ and an effective radius $ r_{eff}>~$~5~arcsec, both in the deeper \textit{g} band. The effective radius criterion was chosen firstly because it is larger than that of the seeing of the observations in DECaLS ($\approx$ 1.5''). In addition, it offers a compromise between sources small enough to have as large a sample as possible (5'' are $\approx$ 500 pc at the NGC~1052 distance of 20 Mpc), but large enough so as not to include a significant fraction of false background projections. Applying the criteria of $r_{eff}> $~5~arcsec and $\mu_g$~(0)~$> $~24~mag~arcsec $ ^{-2}$ and after removing false detections by visual inspection {(process carry out by one member of the team)}, we obtained a sample of 42 LSBGs. We list the final sample in Table \ref{tab:IDs}. We show individual stamp images of each LSBG of the sample in Fig. \ref{fig:Ap1} and we list in Table \ref{tab:params} their structural and photometric properties. For the designation of objects, we follow the guidelines established by the International Astronomical Union\footnote{\url{http://cdsweb.u-strasbg.fr/Dic/iau-spec.html}}. As we use data for general purpose, we opted for the use of the surnames of the authors of this work (RCP), followed by the sequence number of the catalog.

Finally, a recommended procedure is to perform a completeness test, which is the fraction of statistically detected objects based on their structural parameters, typically surface brightness and radius. However, in our case this is unfeasible. In the first place, because the depth of the images is variable throughout the explored area, and so the detectability will vary depending on the depth of the data between different regions (and bands, given that we use the g+r sum image  for detection). Additionally, the LSBG identification procedure includes a visual inspection, {which is} not parameterizable. For these reasons, we do not provide a completeness test, however we do make a comparison with previous works.

\begin{table*}
\centering
\caption{Low-surface-brightness galaxies with $\mu_{g}(0)>$ 24 mag arcsec$^{-2}$ and $r_{eff}~>$~5 arcsec detected in the large-scale environment of NGC~1052. Previously identified objects are noted as identified by: $^a$ \cite{2000A&AS..145..415K}, $^b$ \cite{2003A&A...412...45P}, $^c$ \cite{2007AJ....133..715W}, $^d$ \cite{2018ApJ...868...96C} and $^e$ \cite{2021ApJS..252...18T}.}
{
\label{tab:IDs}
\begin{tabular}{lccl}
\hline
{ID}  &  {RA (\degree)} &  {Dec. (\degree)} &  Previous identification \\
  & {(J2000)} & {(J2000)} & \\
\hline
RCP 1 & 37.2722 & -10.6033 & New \\
RCP 2 & 38.2332 & -9.0575 & New \\
RCP 3 & 38.5704 & -8.9949 & Ta21-11714$^e$ \\
RCP 4 & 39.0332 & -9.9845 & Ta21-11772$^e$ \\
RCP 5 & 39.0430 & -8.3067 & Ta21-11825$^e$ \\
RCP 6 & 39.2677 & -5.2661 & New \\
RCP 7 & 39.4813 & -6.2566 & Ta21-11906$^e$ \\
RCP 8 & 39.5983 & -8.2217 & New \\
RCP 9 & 39.6241 & -7.9257 & NGC~1052-DF7$^d$ \\
RCP 10 & 39.6952 & -6.2366 & Ta21-12088$^e$ \\
RCP 11 & 39.8028 & -8.1408 & NGC~1052-DF5$^d$ \\
RCP 12 & 39.8128 & -8.1160 & NGC~1052-DF4$^d$ \\
RCP 13 & 39.8270 & -7.5369 & New \\
RCP 14 & 39.8617 & -7.3707 & New \\
RCP 15 & 39.9104 & -7.4737 & New \\
RCP 16 & 39.9139 & -8.2285 & Ta21-12000$^e$ \\
RCP 17 & 39.9696 & -8.2121 & New \\
RCP 18 & 40.0194 & -8.4461 & NGC~1052-DF1$^d$ \\
RCP 19 & 40.0341 & -7.9473 & Ta21-12130$^e$ \\
RCP 20 & 40.0820 & -7.9847 & Ta21-12132$^e$ \\
RCP 21 & 40.1200 & -8.2434 & New \\
RCP 22 & 40.1502 & -9.4977 & New \\
RCP 23 & 40.1719 & -11.0829 & New \\
RCP 24 & 40.1894 & -7.6470 & NGC~1052-DF8$^d$; Ta21-12095$^e$ \\
RCP 25 & 40.2277 & -5.3300 & Ta21-12151$^e$ \\
RCP 26 & 40.2897 & -8.2968 & New \\
RCP 27 & 40.3126 & -7.4934 & New \\
RCP 28 & 40.4215 & -8.3475 & New \\
RCP 29 & 40.4451 & -8.4028 & [KKS2000] 04$^a$; LEDA 3097693$^b$; NGC~1052-DF2$^d$; Ta21-12200$^e$ \\
RCP 30 & 40.4475 & -8.7854 & Ta21-12203$^e$ \\
RCP 31 & 40.6033 & -9.4483 & Ta21-12267$^e$ \\
RCP 32 & 40.6202 & -8.3768 & New \\
RCP 33 & 40.6504 & -8.0426 & New \\
RCP 34 & 40.6583 & -7.3381 & WHI B0240-07$^c$ \\
RCP 35 & 40.6963 & -7.7721 & Ta21-12129$^e$ \\
RCP 36 & 40.7636 & -8.0140 & New \\
RCP 37 & 40.8702 & -7.8732 & New \\
RCP 38 & 40.9303 & -6.9219 & Ta21-12315$^e$ \\
RCP 39 & 41.1604 & -7.1764 & New \\
RCP 40 & 41.4845 & -7.6496 & New \\
RCP 41 & 41.8189 & -8.2923 & Ta21-12521$^e$ \\
RCP 42 & 43.2093 & -7.8612 & Ta21-12786$^e$ \\
\hline
\end{tabular}
}
\end{table*}

\begin{table*}
\centering
\caption{Structural and photometric parameters of the sample}
{
\label{tab:params}
\begin{tabular}{lcccccccc}
\hline
{ID}   &  {$r_{eff}$} & {$\mu_{g}(0)$} & {$<\mu_{g}>_{eff}$} & {$n$} & b/a &  {$g$}  &   {$g-r$} &   {$g-z$} \\
  & {[arcsec]} & {[mag arcsec$^{-2}$]} & {[mag arcsec$^{-2}$]} & &  & {[mag]} & {[mag]}  & {[mag]} \\
\hline
RCP 1 &  6.0$\pm$2.1 &          25.9$\pm$0.2 &    27.0$\pm$0.5    & 0.95$\pm$0.55& 0.71$\pm$0.07&  21.17$\pm$0.11 &        0.63$\pm$0.20 & 1.06$\pm$0.26  \\
RCP 2 &  5.4$\pm$0.9 &          26.5$\pm$0.3 &    27.0$\pm$0.5    & 0.51$\pm$0.33& 0.82$\pm$0.13&  21.33$\pm$0.11 &        0.57$\pm$0.22 & 0.37$\pm$0.38  \\
RCP 3 &  8.3$\pm$0.5 &          25.0$\pm$0.1 &    25.8$\pm$0.3    & 0.69$\pm$0.09& 0.81$\pm$0.02&  19.17$\pm$0.02 &        0.58$\pm$0.03 & 0.80$\pm$0.05  \\
RCP 4 &  6.0$\pm$0.4 &          24.2$\pm$0.1 &    25.7$\pm$0.4    & 0.90$\pm$0.16& 0.45$\pm$0.02&  19.82$\pm$0.03 &        0.72$\pm$0.05 & 0.97$\pm$0.08  \\
RCP 5 &  5.2$\pm$0.3 &          24.5$\pm$0.1 &    25.3$\pm$0.4    & 0.66$\pm$0.11& 0.71$\pm$0.03&  19.75$\pm$0.02 &        0.31$\pm$0.04 & 0.42$\pm$0.07  \\
RCP 6 &  9.2$\pm$4.5 &          26.3$\pm$0.2 &    27.2$\pm$0.4    & 0.78$\pm$0.48& 0.86$\pm$0.09&  20.35$\pm$0.14 &        - &     - \\                      
RCP 7 &  10.8 $\pm$0.3 &        24.1$\pm$0.1 &    25.2$\pm$0.2    & 0.90$\pm$0.06& 0.75$\pm$0.01&  18.02$\pm$0.01 &        0.51$\pm$0.02 & 0.64$\pm$0.03  \\
RCP 8 & 5.6 $\pm$2.4 &  26.2$\pm$0.2 &    27.5$\pm$0.5    & 1.05$\pm$0.50& 0.61$\pm$0.11&  21.80$\pm$0.10 &        0.39$\pm$0.22 & 1.07$\pm$0.39  \\
RCP 9 & 12.8 $\pm$1.5 &         25.9$\pm$0.2 &    27.3$\pm$0.2    & 0.71$\pm$0.18& 0.51$\pm$0.03&  19.73$\pm$0.04 &        0.45$\pm$0.08 & 0.81$\pm$0.15  \\
RCP 10 & 8.1 $\pm$1.1 &         24.5$\pm$0.1 &    25.8$\pm$0.3    & 1.14$\pm$0.21& 0.81$\pm$0.03&  19.24$\pm$0.03 &        0.68$\pm$0.05 & - \\          
RCP 11 & 10.0 $\pm$0.8  &       25.9$\pm$0.2 &    26.5$\pm$0.3    & 0.55$\pm$0.12& 0.84$\pm$0.04&  19.51$\pm$0.05 &        0.64$\pm$0.18 & 0.95$\pm$0.15  \\
RCP 12 & 16.8 $\pm$0.4 &        24.3$\pm$0.1 &    25.2$\pm$0.1    & 0.80$\pm$0.04& 0.86$\pm$0.01&  17.07$\pm$0.01 &        0.63$\pm$0.01 & 1.06$\pm$0.02  \\
RCP 13 & 11.0 $\pm$0.7 &        25.4$\pm$0.1 &    26.2$\pm$0.2    & 0.75$\pm$0.12& 0.85$\pm$0.03&  19.02$\pm$0.02 &        0.49$\pm$0.04 & 0.78$\pm$0.06  \\
RCP 14 & 12.3 $\pm$4.1 &        26.9$\pm$0.4 &    27.8$\pm$0.3    & 0.61$\pm$0.35& 0.68$\pm$0.07&  20.36$\pm$0.17 &        - &     - \\                      
RCP 15 & 8.1 $\pm$0.8 &         25.6$\pm$0.1 &    26.8$\pm$0.3    & 0.63$\pm$0.18& 0.50$\pm$0.03&  20.29$\pm$0.04 &        0.56$\pm$0.08 & 0.50$\pm$0.18  \\
RCP 16 & 6.1 $\pm$0.7 &         24.3$\pm$0.1 &    25.6$\pm$0.4    & 0.82$\pm$0.19& 0.56$\pm$0.02&  19.63$\pm$0.02 &        0.55$\pm$0.05 & 0.77$\pm$0.06  \\
RCP 17 & 6.4 $\pm$4.0 &         26.6$\pm$0.3 &    27.9$\pm$0.5    & 1.14$\pm$0.50& 0.53$\pm$0.17&  21.52$\pm$0.16 &        0.84$\pm$0.54 & 1.01$\pm$0.45  \\
RCP 18 & 24.7 $\pm$6.2 &        26.3$\pm$0.2 &    27.9$\pm$0.7    & 1.10$\pm$0.35& 0.72$\pm$0.05&  18.90$\pm$0.65 &        - &-\\                        
RCP 19 & 8.6 $\pm$0.5 &         24.6$\pm$0.1 &    25.7$\pm$0.3    & 0.88$\pm$0.11& 0.75$\pm$0.02&  18.99$\pm$0.02 &        0.49$\pm$0.03 & 0.81$\pm$0.05  \\
RCP 20 & 6.8 $\pm$0.2 &         24.1$\pm$0.1 &    25.0$\pm$0.3    & 0.77$\pm$0.07& 0.77$\pm$0.02&  18.84$\pm$0.01 &        0.55$\pm$0.02 & 0.86$\pm$0.04  \\
RCP 21 & 8.9 $\pm$1.9 &         26.0$\pm$0.2 &    26.9$\pm$0.4    & 0.74$\pm$0.26& 0.79$\pm$0.17&  20.17$\pm$0.07 &        0.55$\pm$0.16 & - \\          
RCP 22 & 5.2 $\pm$1.0 &         25.5$\pm$0.1 &    26.6$\pm$0.4    & 0.59$\pm$0.35& 0.52$\pm$0.06&  21.01$\pm$0.05 &        0.80$\pm$0.12 & 1.29$\pm$0.15  \\
RCP 23 & 9.9 $\pm$1.2 &         25.5$\pm$0.1 &    26.6$\pm$0.3    & 0.86$\pm$0.20& 0.76$\pm$0.03&  19.59$\pm$0.05 &        0.66$\pm$0.08 & 0.95$\pm$0.12  \\
RCP 24 & 7.1 $\pm$0.3 &         25.0$\pm$0.1 &    25.6$\pm$0.3    & 0.58$\pm$0.08& 0.80$\pm$0.03&  19.37$\pm$0.02 &        0.35$\pm$0.04 & 0.35$\pm$0.09  \\
RCP 25 & 6.4 $\pm$0.8 &         24.8$\pm$0.1 &    25.7$\pm$0.4    & 0.87$\pm$0.22& 0.85$\pm$0.04&  19.69$\pm$0.04 &        0.64$\pm$0.06 & 0.96$\pm$0.09  \\
RCP 26 & 9.1 $\pm$0.6 &         24.7$\pm$0.1 &    25.9$\pm$0.3    & 0.77$\pm$0.13& 0.63$\pm$0.02&  19.08$\pm$0.02 &        - &     - \\
RCP 27 & 6.8 $\pm$0.6 &         25.5$\pm$0.1 &    26.5$\pm$0.4    & 0.69$\pm$0.19& 0.66$\pm$0.05&  20.35$\pm$0.07 &        0.57$\pm$0.11 & 0.94$\pm$0.40  \\
RCP 28 & 7.1 $\pm$1.7 &         26.1$\pm$0.2 &    27.0$\pm$0.5    & 0.78$\pm$0.37& 0.74$\pm$0.08&  20.79$\pm$0.17 &        1.05$\pm$0.37 & - \\          
RCP 29 & 21.3 $\pm$0.3 &  24.7$\pm$0.1 &    25.3$\pm$0.1    & 0.58$\pm$0.02& 0.89$\pm$0.01&  16.62$\pm$0.01 &    0.60$\pm$0.01 &     0.94$\pm$0.01  \\
RCP 30 & 7.6 $\pm$0.3 &         24.4$\pm$0.1 &    25.3$\pm$0.3    & 0.71$\pm$0.07& 0.72$\pm$0.02&  18.94$\pm$0.02 &        0.63$\pm$0.03 & 0.86$\pm$0.05  \\
RCP 31 & 6.5 $\pm$0.5 &         24.7$\pm$0.1 &    25.9$\pm$0.4    & 0.74$\pm$0.14& 0.54$\pm$0.02&  19.84$\pm$0.03 &        0.56$\pm$0.06 & 0.95$\pm$0.08  \\
RCP 32 & 23.0 $\pm$7.5 &  27.8$\pm$0.7 &    28.6$\pm$0.5    & 0.50$\pm$0.36& 0.70$\pm$0.07&  19.75$\pm$0.41 &        - &     - \\                      
RCP 33 & 9.9 $\pm$0.5 &         25.3$\pm$0.1 &    26.0$\pm$0.3    & 0.77$\pm$0.10& 0.88$\pm$0.03&  19.06$\pm$0.03 &        0.46$\pm$0.08 & 0.69$\pm$0.18  \\
RCP 34 & 17.8 $\pm$0.1 &        24.1$\pm$0.1 &    24.7$\pm$0.1    & 0.46$\pm$0.01& 0.82$\pm$0.01&  16.41$\pm$0.01 &        0.35$\pm$0.01 & 0.46$\pm$0.01  \\
RCP 35 & 8.6 $\pm$0.3 &         24.3$\pm$0.1 &    25.5$\pm$0.3    & 0.82$\pm$0.08& 0.61$\pm$0.01&  18.85$\pm$0.01 &        0.50$\pm$0.03 & 0.80$\pm$0.04  \\
RCP 36 & 6.7 $\pm$0.6 &         25.5$\pm$0.1 &    26.5$\pm$0.3    & 0.45$\pm$0.15& 0.49$\pm$0.03&  20.39$\pm$0.02 &        0.41$\pm$0.06 & 0.62$\pm$0.11  \\
RCP 37 & 7.0 $\pm$0.8 &         25.9$\pm$0.2 &    27.0$\pm$0.3    & 0.60$\pm$0.23& 0.57$\pm$0.05&  20.75$\pm$0.03 &        0.72$\pm$0.07 & 1.22$\pm$0.12  \\
RCP 38 & 8.4 $\pm$0.6 &         25.5$\pm$0.1 &    26.2$\pm$0.3    & 0.61$\pm$0.12& 0.82$\pm$0.03&  19.57$\pm$0.03 &        0.58$\pm$0.05 & 1.16$\pm$0.08  \\
RCP 39 & 11.2 $\pm$0.3 &  24.1$\pm$0.1 &    25.6$\pm$0.2    & 0.79$\pm$0.05& 0.45$\pm$0.01&  18.37$\pm$0.01 &        0.42$\pm$0.03 & 0.61$\pm$0.04  \\
RCP 40 & 6.4 $\pm$0.3 &         24.1$\pm$0.1 &    25.2$\pm$0.3    & 0.82$\pm$0.10& 0.70$\pm$0.03&  19.16$\pm$0.01 &        0.42$\pm$0.02 & 0.65$\pm$0.04  \\
RCP 41 & 5.5 $\pm$0.3 &         24.5$\pm$0.1 &    25.5$\pm$0.4    & 0.75$\pm$0.12& 0.72$\pm$0.03&  19.76$\pm$0.02 &        0.46$\pm$0.03 & 0.76$\pm$0.06  \\
RCP 42 & 5.9 $\pm$1.1 &         25.3$\pm$0.1 &    26.6$\pm$0.4    & 0.91$\pm$0.36& 0.49$\pm$0.05&  20.80$\pm$0.05 &        0.54$\pm$0.12 & 0.97$\pm$0.16  \\
\hline
\end{tabular}
}
\end{table*}

\subsection{Comparison with previous studies}

Recent works explored the presence of LSBGs in the environment of the NGC~1052 group using deep imaging. First, \cite{2018ApJ...868...96C} (hereafter Co18), explored the presence of LSBGs in a region of $\approx$ 3$\times$3 degrees centered around NGC~1052 (see Section \ref{sec:data} for a comparative description of these data). Through a process of visual identification, {Co18} reported the presence of six LSBGs without specific selection criteria that were later observed with the Hubble Space Telescope producing high-resolution imaging. We identified these six galaxies  in our work; see Table \ref{tab:IDs}. The structural and morphological properties reported by Co18 in these six LSBGs are equivalent to those reported in our work. However we find an exception in the case of RCP~9 (NGC~1052-DF7): we report an effective radius of 12.8 arcsec, while Co18 provide a value of 18.9 arcsec. Regarding the properties of the six galaxies in common with Co18 and those unidentified, we did not find significant differences for the central surface brightness with a mean of $\overline{\mu_{g}(0)}$~=~25.3~mag~arcsec$^{-2}$ for the galaxies in common and $\overline{\mu_{g}(0)}$~=~25.2~mag~arcsec$^{-2}$ for those not identified by Co18. However, we do find significant differences in magnitude and effective radius, with \textit{$\overline{g}$} = 18.5 mag and $\overline{r_{eff}}$~=~15.5~arcsec for those galaxies in common with Co18, and \textit{$\overline{g}$} = 19.8 mag and $\overline{r_{eff}}$~=~8.5~arcsec for those not identified by Co18. These could be related to the higher resolution and depth of our data compared to those used by Co18. Additionally, they could be related to the Co18 identification method, which is not systematic with certain criteria, but visual, selecting interesting targets for a high-resolution follow-up by the Hubble Space Telescope.

An interesting comparison can {also} be made with the work by \cite{2021ApJS..252...18T} (hereafter Ta21), {who} explored the presence of LSBGs over $\sim$5000 deg$^2$ from the first three years of imaging data from the Dark Energy Survey (DES), including the region of the NGC~1052 environment that we explore here. The criteria to define a LSBG according to Ta21 are $r_{eff}> $~2.5~arcsec and <$\mu_{g}$>$_{eff}>$~24.2~mag~arcsec$^{-2}$, therefore less restrictive {in terms of effective radius} than our LSBG criteria. We find that 17 of the 42 galaxies identified in our work appear in the Ta21 catalog; see Table \ref{tab:IDs}. The structural and photometric properties reported by Ta21 on these 17 galaxies in common are similar to what we report here. However, we find a small difference for the central surface brightness, in which the values in our work are on average 0.15 mag arcsec$^{-2}$ brighter than those found by Ta21. Given that magnitude, effective radius, and S\'ersic index are similar between both catalogs, we attribute this difference to the PSF deconvolution process as part of the S\'ersic model fit, which was not carried out by Ta21. We find significant differences in average parameters between galaxies in common and those not detected by Ta21 in our sample. The LSBGs in our catalog not detected by Ta21 are on average fainter ($\overline{g}$~=~19.8~vs.~19.2~mag), larger ($\overline{r_{eff}}$~=~10.4~vs.~8.1~arcsec), and have a significantly lower surface brightness ($\overline{\mu_{g}(0)}$~=~25.7~vs.~24.6~mag~arcsec$^{-2}$) than those LSBGs in our catalog in common with Ta21. In turn, there are ten galaxies with the criteria of $r_{eff}$~>~5 arcsec and $\mu_{g}(0)$~>~24.0~mag~arcsec$^{-2}$ that are found in the Ta21 catalog but not in ours. Eight of these ten galaxies were preliminary identified in our identification process with parameters very close to the selection criteria but not fulfilling it: four were discarded for having $r_{eff}$~<~5~arcsec and four for having $\mu_{g}(0)$~>~24.0~mag~arcsec$^{-2}$. Of the two remaining objects, we considered one of them to have multiple components, and so it would have been discarded in the visual inspection criteria (\texttt{object\_id} = 11784 in Ta21; R.A. = 38.6509, Dec. = -10.9843) and another object that was undetected by our procedure (\texttt{object\_id} = 2752 in Ta21; R.A. = 37.2745, Dec. = -10.6457). 

\subsection{Properties of the low-surface-brightness galaxies}\label{sec:Properties}

\begin{figure*}
\centering
        \includegraphics[width=1.0\textwidth]{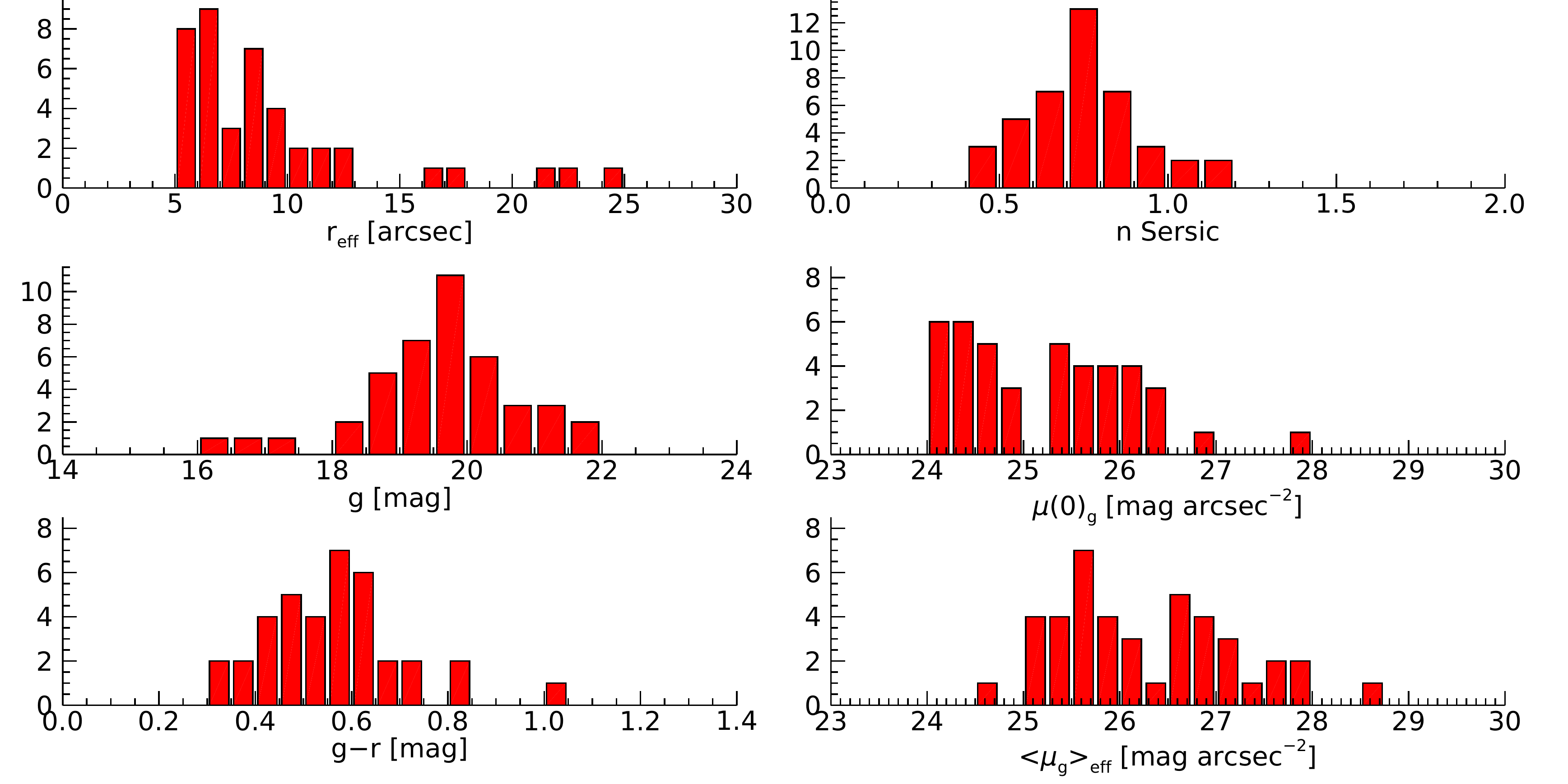}
    \caption{Histograms of structural and photometric properties of the sample.}
    \label{fig:Stats}
\end{figure*}

Figure \ref{fig:Stats} shows the distributions of the most relevant structural and photometric properties of the LSBG sample. The S\'ersic indexes show a clear distribution centered on n~=~0.77 with a standard deviation of $\sigma$~=~0.22, with only four objects exceeding n~=~1. This distribution in S\'ersic index is compatible with a sample whose criterion is the central surface brightness \citep{2015ApJ...807L...2K, 2017MNRAS.468..703R}. In turn, the surface brightness distributions show behaviors consistent with the selection criteria. The central surface brightness shows an approximately constant distribution up to approximately $\mu_{g}(0)$~>~26.5~mag~arcsec$^{-2}$. {Two objects have} extremely low surface brightness (RCP~14 and RCP~32). The distribution is more homogeneous in the case of the average surface brightness within the effective radius, as expected, with galaxies ranging <$\mu_{g}$>$_{eff}$~$=$~25-28~mag~arcsec $ ^{-2}$ with the two previously mentioned objects {having} <$\mu_{g}$>$_{eff}$~>~28~mag~arcsec $ ^{-2}$. Regarding effective radius, we find a compact distribution up to 15 arcsec. Above this value we find a number of galaxies that have an effective radius larger than the continuity of the distribution. The apparent magnitude distribution in the \textit{g} band {is} well centered at a value of \textit{g}~=~19.6 mag, with a standard deviation of $\sigma$~=~1.2 mag. Finally, the analysis of the \textit{g-r} color (we ruled out the use of the \textit{z} band because of its shallow depth) has the potential of showing the presence of sources projected in the background, which are reddened and are not physically associated with the structure of interest in the analysis, in our case the environment of the group of galaxies of NGC~1052. While the histogram of the \textit{g-r} color {shows} objects that have a value higher than \textit{g-r}~=~0.7, taking into account the photometric errors listed in Table \ref{tab:params}, all LSBGs in the sample have a \textit{g-r} compatible with \textit{g-r}~<~0.7 mag within photometric errors. Therefore, no object can be considered a projection in the background based on the photometric colors (however, we note that not all objects have a \textit{g-r} color value because of the lower S/N in the \textit{r} band compared to the \textit{g} band; see Table \ref{tab:params}). This is expected because the criterion of r$_{eff}$~<~5~arcsec is effective in selecting very close objects that are therefore not reddened. {Due to the large photometric uncertainties in these extremely low-surface-brightness objects, we ruled out the use of photo-z in the analysis.}

\section{Analysis}

\subsection{Potential GCs in the LSBGs}

\begin{figure*}
\centering
        \includegraphics[width=1.0\textwidth]{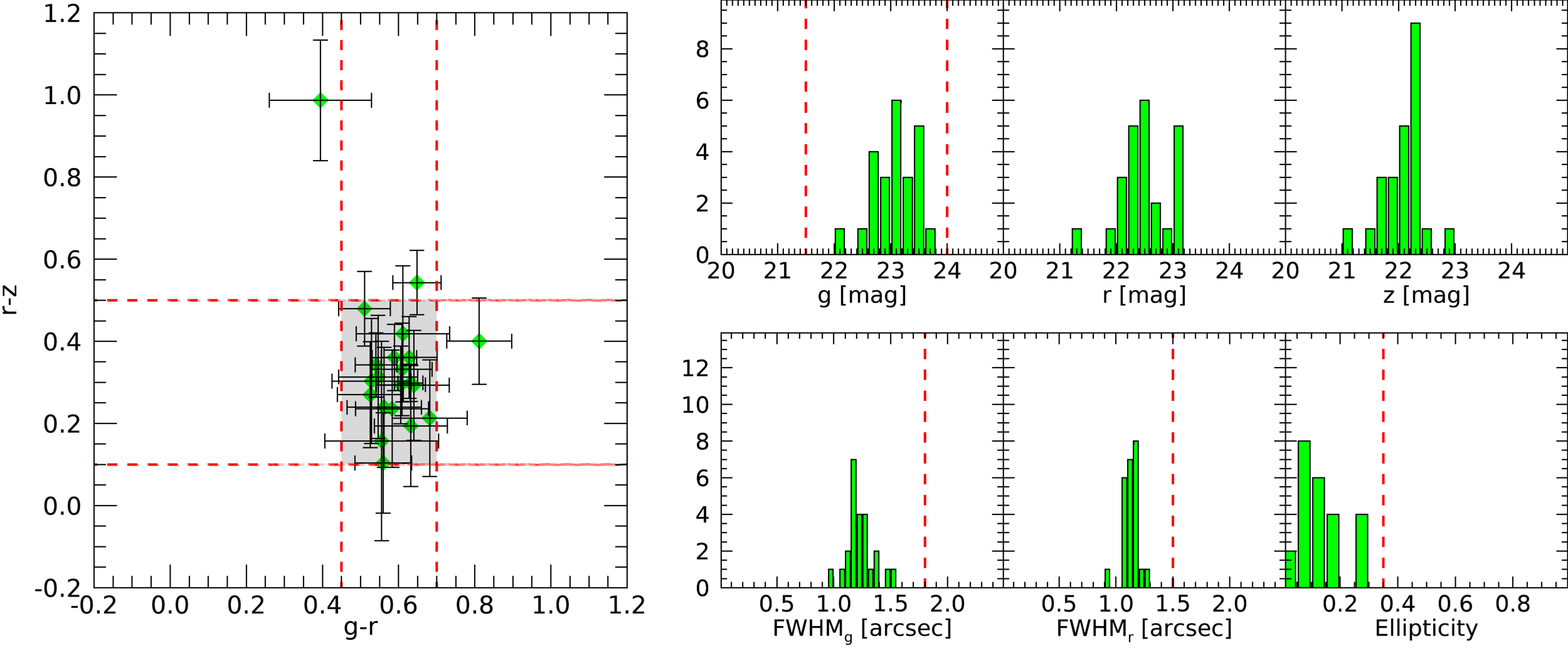}
    \caption{Properties of spectroscopically confirmed {GCs} by previous works in the galaxies RCP~12 (NGC~1052-DF4) and RCP~29 (NGC~1052-DF2) in our data. The left panel shows the colors of the GCs in the \textit{g-r} vs. \textit{r-z} map. The right panels show histograms with their photometric and structural properties. The dashed red lines show the criteria for the detection of new GC candidates for the galaxies in our sample.}
    \label{fig:GCs_props}
\end{figure*}

The analysis of GCs is of great interest in the study of LSBGs as they have the potential to provide relatively accurate distance estimations \citep[e.g.,][]{2012Ap&SS.341..195R} and are tracers of the dynamical mass of the host galaxy once their radial velocities have been calculated \citep[e.g.,][]{2016ApJ...819L..20B}. We carried out a statistical analysis for the presence of GC systems around the detected LSBGs. The main difficulty in detecting GCs is distinguishing them from point sources both in background and foreground. The use of high-resolution, multi-band photometric data is helpful in reducing the interloper fraction. In our case, the ground-based resolution with an average seeing of approximately 1 arcsec and the availability of the \textit{g}, \textit{r,} and \textit{z} {bands} allows a significant reduction in the fraction of false detections, although not total. Therefore, in the absence of spectroscopic measurements, only a statistical analysis is feasible with the available data. In this situation, an over-density of sources compatible with GCs over the area where the LSBG is located would indicate the presence of a GC system, but no individual sources can be confidently claimed to be GCs.

In order to constrain the properties of potential GCs in the LSBGs of our sample, we used the spectroscopically confirmed GCs of RCP~12 (NGC1052-DF4) and RCP~29 ([KKS2000] 04, NGC1052-DF2) provided by { \citet[][]{2018ApJ...856L..30V}, \citet[][]{2019A&A...625A..76E}, \citet[][]{2019ApJ...874L...5V} and \citet[][]{2021ApJ...909..179S}}, with a total of 24 GCs. The photometric and structural parameters of these confirmed GCs in our data will provide a template for selecting GC candidates in other galaxies. 

Characterization of the GCs was carried out by performing aperture photometry using \texttt{SExtractor} in dual mode, with the \textit{g+r} sum image {for} detection {and} the individual \textit{g}, \textit{r,} and \textit{z} bands for photometry. This is done on individual stamps of 4 arcmin centered on the LSBGs in the catalog. For the photometry of point sources, we selected an aperture of 12 pixels or 3.24 arcsec (approximately 3$\times$FWHM or seeing of the data) {and} a detection threshold of 5$\sigma$ with the aim of selecting high-S/N sources. The resulting parameters for the confirmed GCs of RCP~12 and RCP~29 are shown in Fig. \ref{fig:GCs_props}. Of the 24 spectroscopically confirmed GCs in the sample, we detected 21 in our data. As can be seen, the GCs have typical colors that are well constrained in the \textit{g-r} versus \textit{r-z} map with an average color of \textit{g-r} = 0.6 mag and \textit{r-z} = 0.3 mag. However the fainter ones appear with larger photometric errors, and in some cases outside this main region in the \textit{g-r} versus \textit{r-z} map. The structural properties also appear well limited. The FWHMs of the GCs are equivalent to the seeing of the data, around 1.2 arcsec in the \textit{g} band and 1.1 arcsec in the \textit{r} band, {and they} therefore appear as point-like or unresolved sources in our ground-based data. The ellipticity of the GCs remains below 0.35. All of these properties are compatible with those found in previous works using high-resolution space data from the Hubble Space Telescope.

We used the photometric and structural ranges in which the spectroscopically confirmed GCs are located to delimit the parameters of potential GCs in the LSBGs of our catalog. In particular, we selected a color range of 0.45~<~\textit{g-r}~<~0.7, 0.1~<~\textit{r-z}~<~0.5, indicated in Fig. \ref{fig:GCs_props} (left panel) by the red dashed lines and the shaded area. We limited the FWHM in the \textit{g} band to 1.8 arcsec and in \textit{r} band to 1.5 arcsec to select point-like sources {while} taking into account possible seeing variations throughout the large area explored. It is expected that different regions were observed at different epochs, and therefore with different seeing. Finally, we limited the ellipticity of the sources to 0.35. We performed aperture photometry in the stamps where the S\'ersic model of the LSBGs has been subtracted, {aiming} to obtain the cleanest possible point source photometry.

As discussed above, a GC system would appear as an overdensity of GC candidates on the LSBGs with respect to a statistical background of interlopers. This is analyzed in two complementary ways. First, we plot in Fig. \ref{fig:GCs_panel} the number of GC candidates per unit area in bins of annular aperture centered on each of the LSBGs of our sample. This is done using 0.45 arcmin {step} annular apertures in the same way for all LSBGs. Additionally, given that there are significant differences in the sizes of the LSBGs, we created a parameter, which we call $\beta$, that takes into account the effective radii of the LSBGs to calculate a possible overdensity located within the value of three times the effective radius for each galaxy, which we define as:

\begin{figure*}
\centering
        \includegraphics[width=0.85\textwidth]{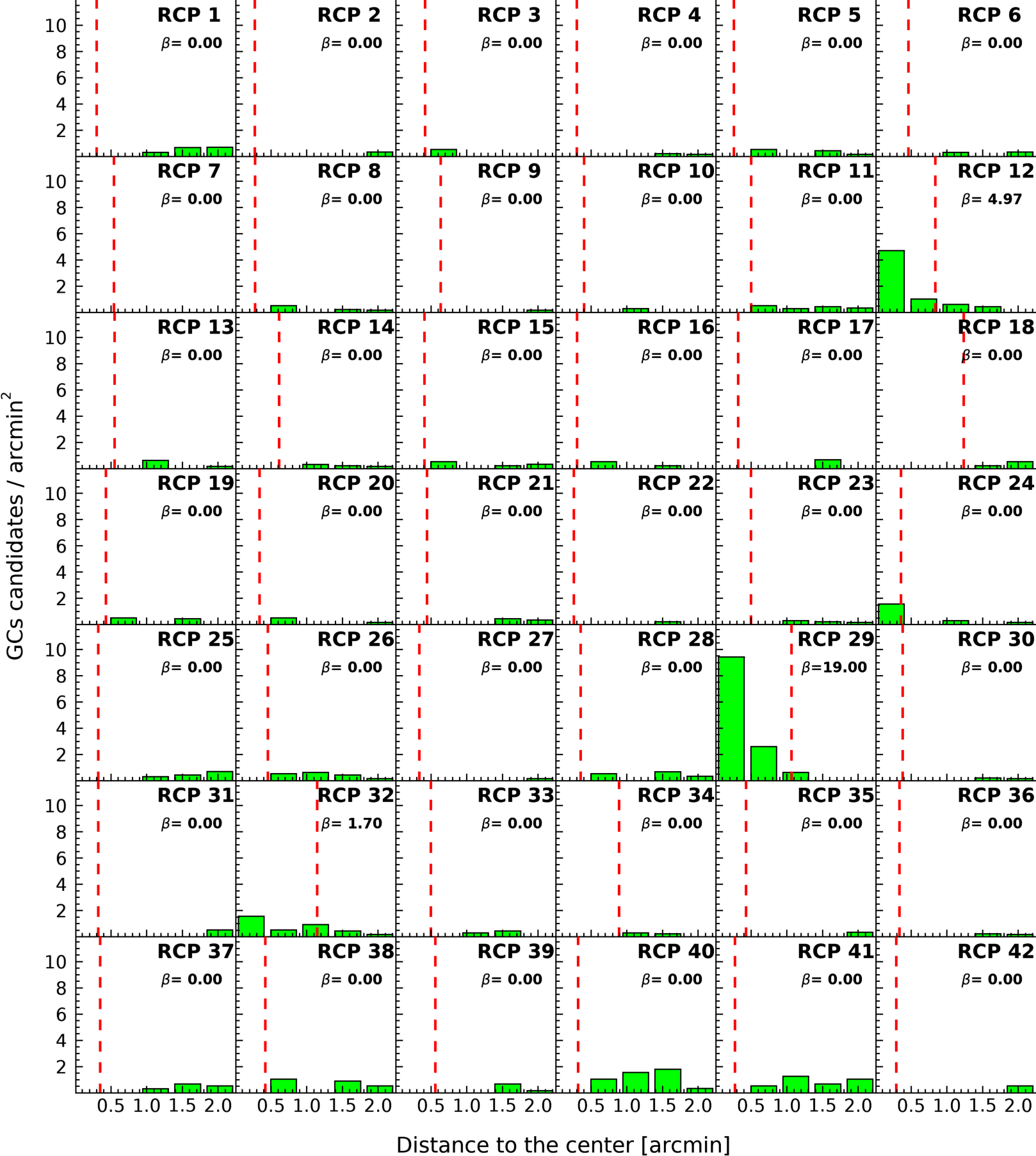}
    \caption{Number of GC candidates per unit area in circular profiles centered on the LSBGs of our sample {(see text)}. The dashed red line shows the 3$\times$r$_{eff}$ value for each galaxy. The $\beta$ value is shown for each galaxy (see main text).}
    \label{fig:GCs_panel}
\end{figure*}

\begin{equation}
    \beta=\frac{GC_{cand}[R<3\times r_{eff}]}{GC_{cand}[R>3\times r_{eff}]}.
\end{equation}

This is the fraction of GC candidates per unit area within the 3$\times$r$_{eff}$ region divided by the number of candidates {per unit area} outside of this region. We used 3$\times$r$_{eff}$ as a value in which most of the GCs are expected to be located, if they exist, motivated by the {works by \cite{2017MNRAS.472L.104F} and \cite{2021MNRAS.502.5921S}}. With this $\beta$ parameter we take into account the size of each LSBG to obtain a possible overdensity of GC candidates on it. It is worth noting that according to this criterion, to claim a detection ($\beta$~>~1), at least one GC candidate must be found within 3$\times$r$_{eff}$, and additionally, the number of GC candidates per unit area must be higher within 3$\times$r$_{eff}$. When no GC candidate is detected within 3$\times$r$_{eff}$, $\beta$ = 0 is obtained, and we {conclude} the absence of a GC system. In the case of obtaining a $\beta$ value close to 1, this is compatible with a background of false detections.

The results of this analysis are the clear detection of a high density of GC candidates in the galaxies RCP~12 and RCP~29, with $\beta$ values well above unity, which is expected. We also detected an overdensity of GC candidates in RCP~32, with $\beta$ = 1.7. These three LSBGs are the only ones in which a nonzero $\beta$ value is obtained. The $\beta$ value in RCP~32 is significantly lower than in RCP~12 and RCP~29, indicating fewer potential GCs. The profile of GC candidates per unit area in RCP~32 is similar to that of RCP~12 and RCP~29, with a decreasing number as the radius grows, from the central region to a distance of  approximately five effective radii. {This confirms} that the overdensity of GC candidates over RCP~32 behaves reliably.

In this analysis we can also find LSBGs {with $\beta$~=~0 but with} a considerable number of GC candidates at larger distances. After analyzing these particular cases in detail, we found that some are pure statistical fluctuations, however we also found that for some LSBGs that are very close to other massive galaxies, the GC candidates belong to the adjacent massive galaxy. Clear examples of this situation are RCP~26, located very close to NGC~1052, and RCP~40, located on the periphery of NGC 1084. Undoubtedly, the GCs belonging to these massive galaxies are clearly detectable with our criteria, having been studied in previous works as in the case of NGC~1052 \citep[][]{2005MNRAS.358..419P}. 

\begin{figure*}
\centering
        \includegraphics[width=1.0\textwidth]{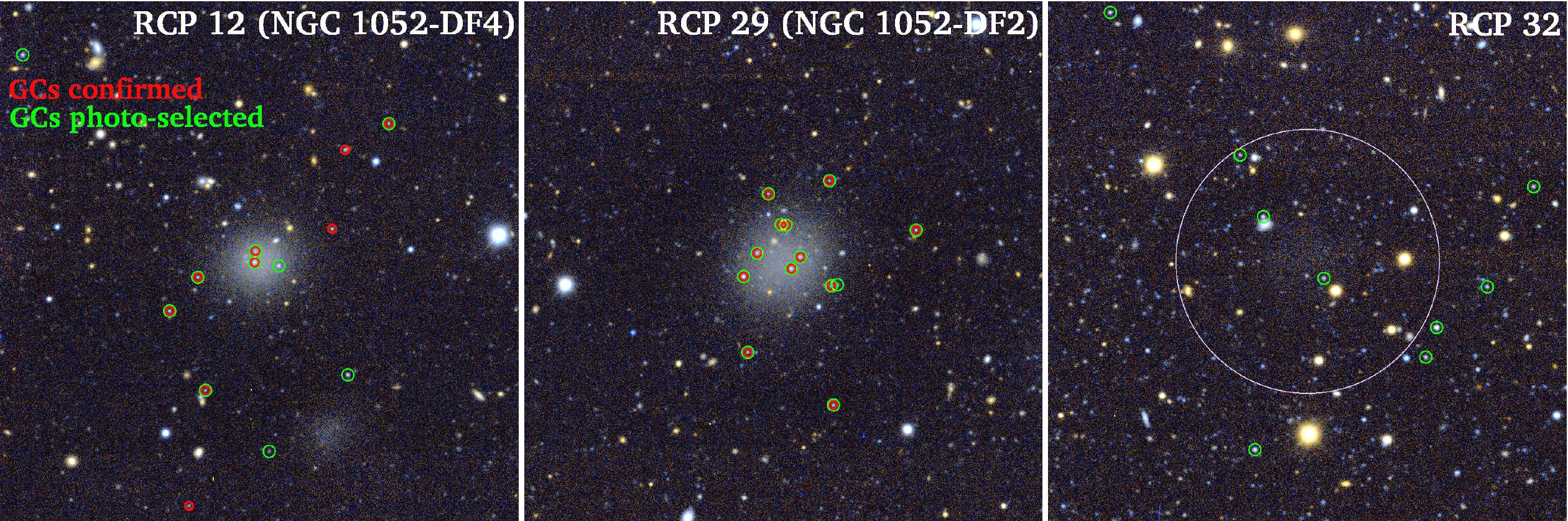}
    \caption{Color-composed stamps using the \textit{g} and \textit{r} bands centered on the three low-surface-brightness galaxies in which we detect a GC system. Stamps are 4 arcmin on a side. The spectroscopically confirmed GCs in RCP~12 (NGC~1052-DF4) and RCP~29 (NGC~1052-DF2) that are detectable in our data are marked with red circles. The GCs that appear after applying our photometric criteria are marked by green circles. We place a purple circle marking 3$\times$r$_{eff}$ on RCP~32, which is barely visible in this color stamp.}
    \label{fig:GCs_gals}
\end{figure*}

According to our analysis, the only GC systems detected in our sample are the already known ones in RCP~12 and RCP~29, and the marginal detection of a GC system in RCP~32. In Fig. \ref{fig:GCs_gals} we show color stamps of these galaxies along with the spectroscopically confirmed GCs by previous works that are detected in our data (in red). Additionally, we show the GC candidates that appear after applying the photometric criteria discussed above (in green). As can be seen in RCP~12 and RCP~29, despite only having ground-based observations in \textit{g}, \textit{r,} and \textit{z} bands, the GC candidates trace the distribution of spectroscopically confirmed GCs remarkably well, with a low fraction of false detections. This efficient filtering of GCs against false detections suggests that we can consider the GC system detected in RCP~32 to be real.

We note that the photometric criteria used for the detection of GC candidates are those corresponding to the specific GC population of RCP~12 and RCP~29 galaxies, and similar between both. Therefore, it is possible that the stellar populations in the case of RCP~32, or another LSBG in the sample, were significantly different. Hence the criteria based on the already detected populations of GCs in RCP~12 and RCP~29 may not be as effective in other galaxies in the sample. Data of higher resolution and with the inclusion of the important \textit{u} band could be more decisive in the search for GCs in the LSBGs.

One of the striking properties of RCP~12 and RCP~29 galaxies, and a source of controversy, is that if they were located at the distance of NGC~1052, at 20 Mpc, the GCs would be more luminous \citep[e.g.,][]{2018Natur.555..629V, 2021ApJ...909..179S} than expected given the universality of the GCLF in galaxies \citep[][]{2012Ap&SS.341..195R}. It is therefore worth exploring whether this also occurs in RCP~32. First, we measured the peak of the luminosity function of the spectroscopically confirmed GCs in our data. These values calculated as the average value of the detected GCs are: $\mu_g$~=~23.16~mag, $\mu_r$~=~22.52~mag, and $\mu_z$~=~22.05~mag for RCP~12 and $\mu_g$~=~23.01~mag, $\mu_r$~=~22.48~mag, and $\mu_z$~=~22.21~mag for RCP~29. We note here that the peak calculated in our data is slightly lower than the equivalent of previous works because we are {missing} a few GCs that are not detected {because of} the point source detection limit of our data, which is lower than that of the Hubble Space Telescope used in previous works. However, it is useful for direct comparison in RCP~32: $\mu_g$ = 23.15 mag, $\mu_r$ = 22.52 mag, and $\mu_z$ = 22.35 mag. We can verify that the GC candidates of RCP~32 have a peak in their luminosity function equivalent to those of RCP~12 and RCP~29, {suggesting that it would be at a similar distance, around 13 Mpc, if the GCLF peak method applies in these galaxies \citep[see][]{2019MNRAS.486.1192T}}. As the GCs of RCP~32 are not spectroscopically confirmed, a certain presence of false GCs is expected. This could have a significant impact given the low number of GC candidates for the calculation of the peak of the GCLF in RCP~32.

\subsection{Spatial correlation with spectroscopic line-of-sight structures}\label{sec:4.1}

An intrinsic problem with very low-surface-brightness sources is the systematic absence of spectroscopic measurements with which to obtain their distances. It is therefore necessary to include in any environmental analysis the possibility of false projections in the LSBG sample, and therefore we need to study spatial structures at different distances in the projected line of sight. To characterize the structures present in the analyzed field, we used the NASA/IPAC Extragalactic Database (NED) to obtain spectroscopic measurements available in the region. We carried out a search in a 8$\times$8 deg region, 1 deg wider on each side than the LSBG detection mosaic in order to trace possible structures that are adjacent to the LSBGs, but outside the boundaries where detection of LSBGs was carried out. This region is therefore 36.37º~<~R.A.~<~44.27º, -12.26º~<~Dec.~<~-4.26º. We used  the search by parameters in NED for this task, unselecting spectroscopy of HII regions or different components of galaxies. 

\begin{figure}
\centering
        \includegraphics[width=\columnwidth]{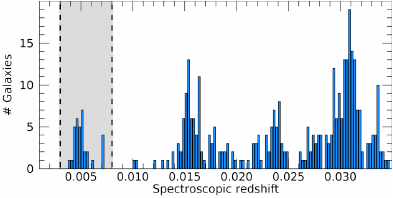}
    \caption{Histogram of galaxies with spectroscopic redshift in the analyzed region. The gray shaded stripe marks the structure of interest associated with the NGC~1052 group of galaxies at 0.003~<~z~<~0.008, corresponding to $\approx$ 900 < V$_{rad}$ < 2400 km s$^{-1}$.}
    \label{fig:Histogram}
\end{figure}

In Fig. \ref{fig:Histogram} we show the histogram of galactic redshifts that we obtained from the NED database in the  selected area described above. There is a peak with a considerable number of galaxies (35 objects, including RCP~29 or NGC~1052-DF2, see Table~\ref{tab:IDs}) with a similar radial velocity to that of NGC~1052 (z~=~0.005). We  mark a gray region  in the histogram that can be considered the redshift interval associated with the environment of NGC~1052 (0.003~<~z~<~0.008), and therefore the region of interest in this work. We note the existence of a virialized-looking structure of 31 galaxies in the interval 0.004~<~z~<~0.006 with a calculated velocity dispersion of $\sigma$~=~135~km~s$^{-1}$ whose mean radial velocity is 1453 km s$^{-1}$, very similar to the radial velocity of NGC~1052 of 1510 km s$^{-1}$. We 
also find a subgroup at approximately z = 0.007 with a velocity dispersion of $\sigma$~=~12~km~s$^{-1}$ centered at a radial velocity of 2114 km s$^{-1}$. Table \ref{tab:specs} provides the name, coordinates, and radial velocities of all the galaxies with spectroscopy in this region. The NGC~1052 environment is considerably isolated in redshift space. We find a void region at 0.007~<~z~<~0.013, finding the first clearly virialized background structure at z~=~0.015, and a more massive structure at z~=~0.031. This isolation in redshift space, along with being the first structure in the foreground, ensures a low fraction of interlopers or false background projections for the LSBGs in the NGC~1052 environment. Nevertheless, it is interesting to explore the spatial correlation of the detected LSBGs with the spectroscopic galaxies in different redshift slices in order to draw their spatial correlation.

\begin{figure*}
\centering
        \includegraphics[width=0.85\textwidth]{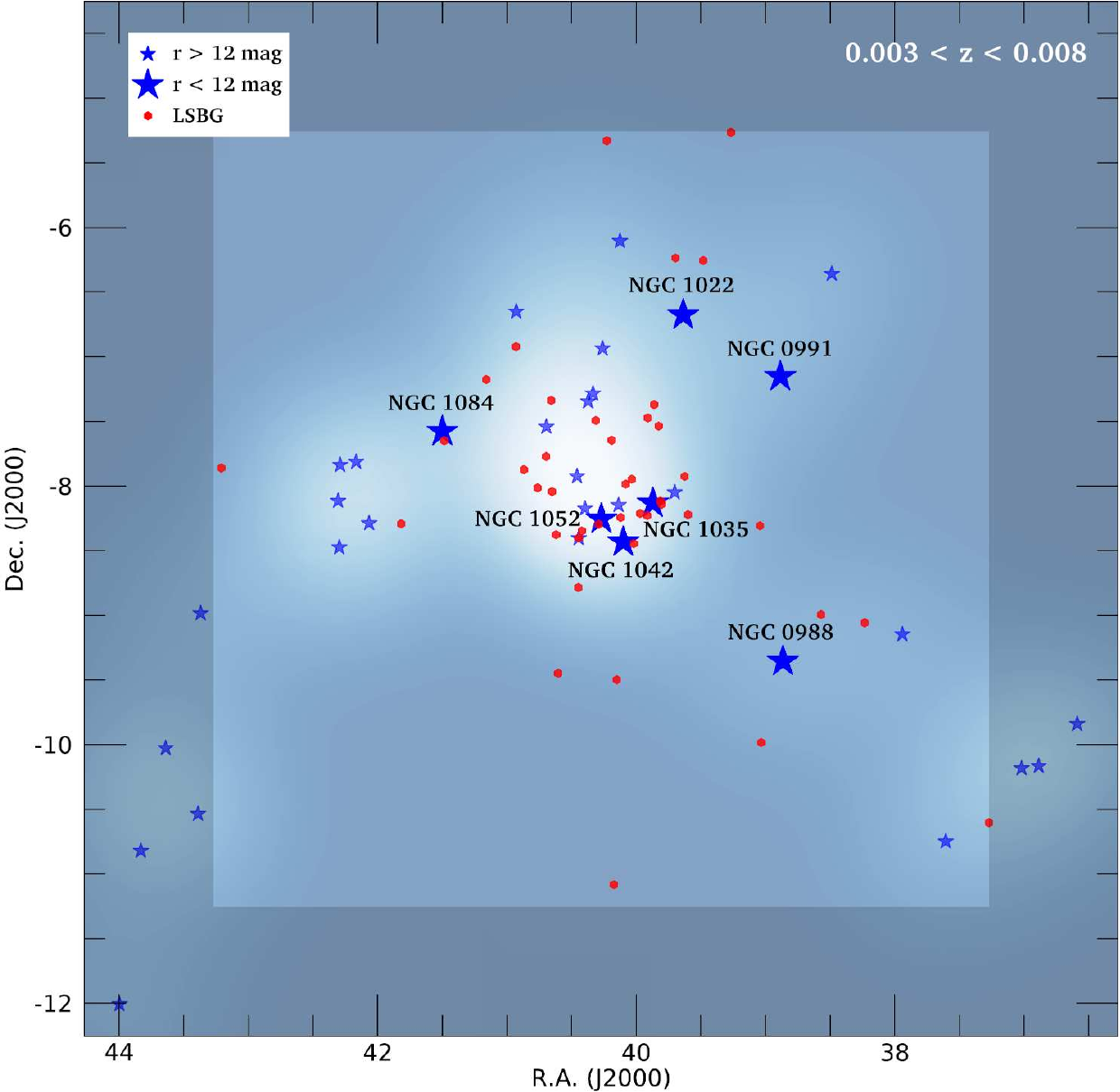}
    \caption{Spatial correlation of the LSBGs (red dots) together with galaxies with spectroscopic measurements (blue stars) in the interval 0.003~<~z~<~0.008 to which the NGC~1052 environment belongs. Galaxies with spectroscopy {are} separated into two groups according to whether they have a magnitude greater or less than \textit{r} = 12 mag. The shaded region marks the limit where an identification of LSBGs has not been carried out but galaxies with spectroscopy are included, for better characterization of the large-scale structure. A white background traces the density of galaxies with spectroscopy for ease of visualization.}
    \label{fig:Spatial_distribution}
\end{figure*}

In Fig. \ref{fig:Spatial_distribution} we plot the spatial distribution of the detected LSBGs with the galaxies with spectroscopy located in the interval 0.003~<~z~<~0.008. We differentiate between galaxies with a magnitude higher or lower than \textit{r} = 12 mag, which is useful to identify the most massive or dominant galaxies in the region. There is a well-defined large-scale structure centered in the analyzed area. We place the names of these dominant galaxies in Fig. \ref{fig:Spatial_distribution}. In the center of this structure are three galaxies with \textit{r} < 12 mag: NGC~1052, NGC~1042, and NGC~1035.  The highest concentration of LSBGs is found around these three dominant galaxies. Surrounding this denser central regions are a number of spectroscopic galaxies, including some massive galaxies with \textit{r} < 12 mag. However, the densities of both spectroscopically confirmed galaxies and LSBGs are low in these outer regions. We can also find regions with {a} total absence of galaxies. In general, we can confirm the good spatial correlation of the LSBGs with the galaxies located in the redshift interval associated with the NGC~1052 environment.

\begin{figure*}
\centering
        \includegraphics[width=1.0\textwidth]{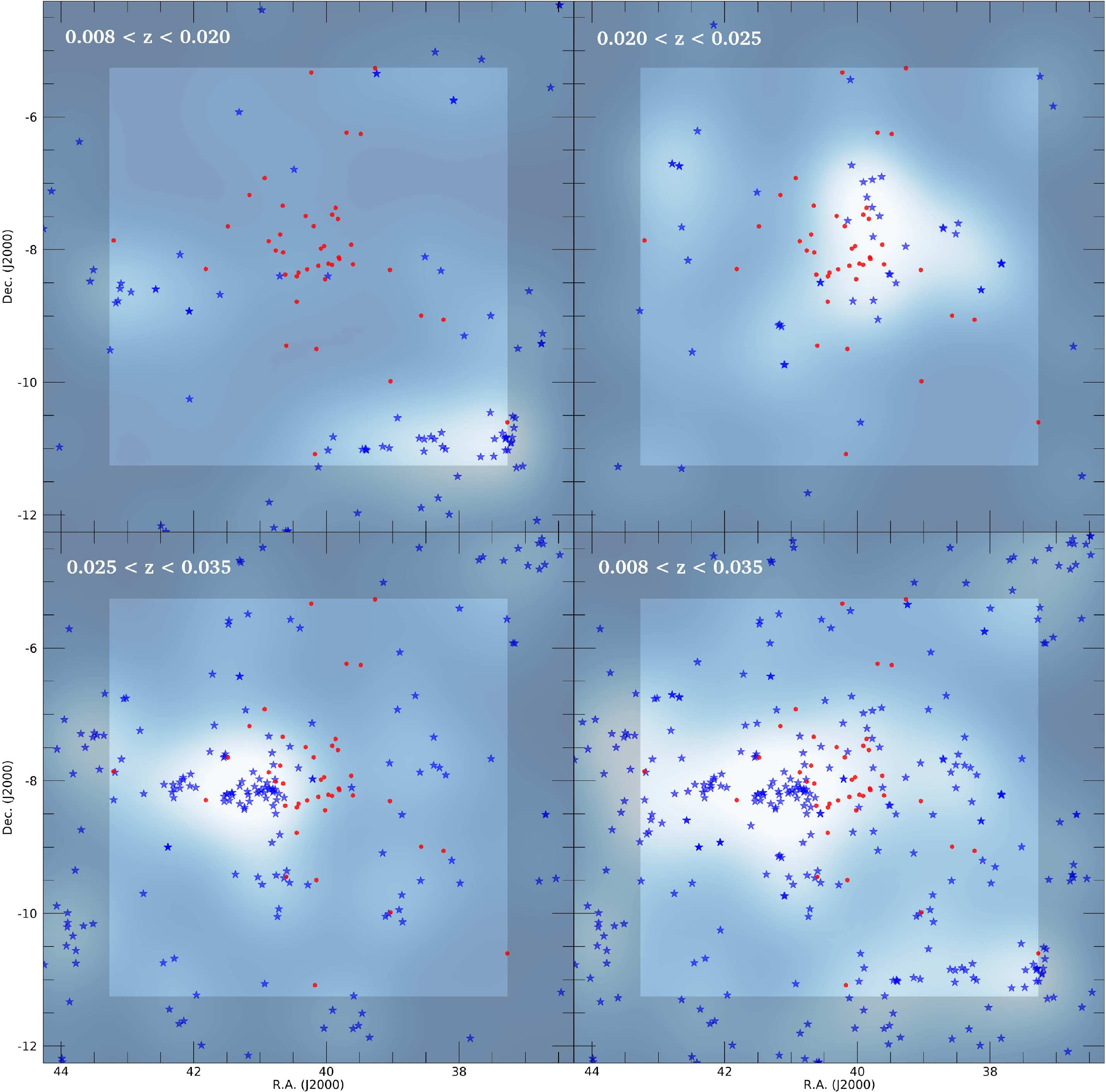}
    \caption{Similar plot to that of Fig. \ref{fig:Spatial_distribution} but in different redshift intervals corresponding to overdensities in the background of the NGC~1052 environment, for each panel. We note that the bottom right panel (0.008 < z < 0.035) is the cumulative of the other three panels.}
    \label{fig:Spatial_distribution_backs}
\end{figure*}

For comparison, we carried out the same analysis, this time with galaxies in the background of the environment of NGC~1052 (0.008~<~z~<~0.035), which can be visualized in Fig. \ref{fig:Spatial_distribution_backs}. We included three maps similar to that of Fig. \ref{fig:Spatial_distribution} corresponding to the overdensities observed at 0.008~< z~< 0.020, 0.020~<~z~<~0.025, and 0.025~<~z~<~0.035 separately, and a fourth panel which corresponds to the cumulative of the three (0.008 < z < 0.035). Visual inspection of these maps allows us to verify that the spatial correlation between LSBGs and galaxies in the redshift range in which NGC~1052 is found is higher than the spatial correlation with background structures. It is interesting to note that the probability that a LSBG belongs to a structure in the background decreases with its distance, because the number density of LSBGs with a given radius decreases with radius \citep{2015ApJ...807L...2K, 2017MNRAS.468..703R}. For this reason, a cut like the criterion of r$_{eff}$~>~5 arcsec that we use in our work tends to select the more nearby objects.

\begin{figure*}
\centering
        \includegraphics[width=1.\columnwidth]{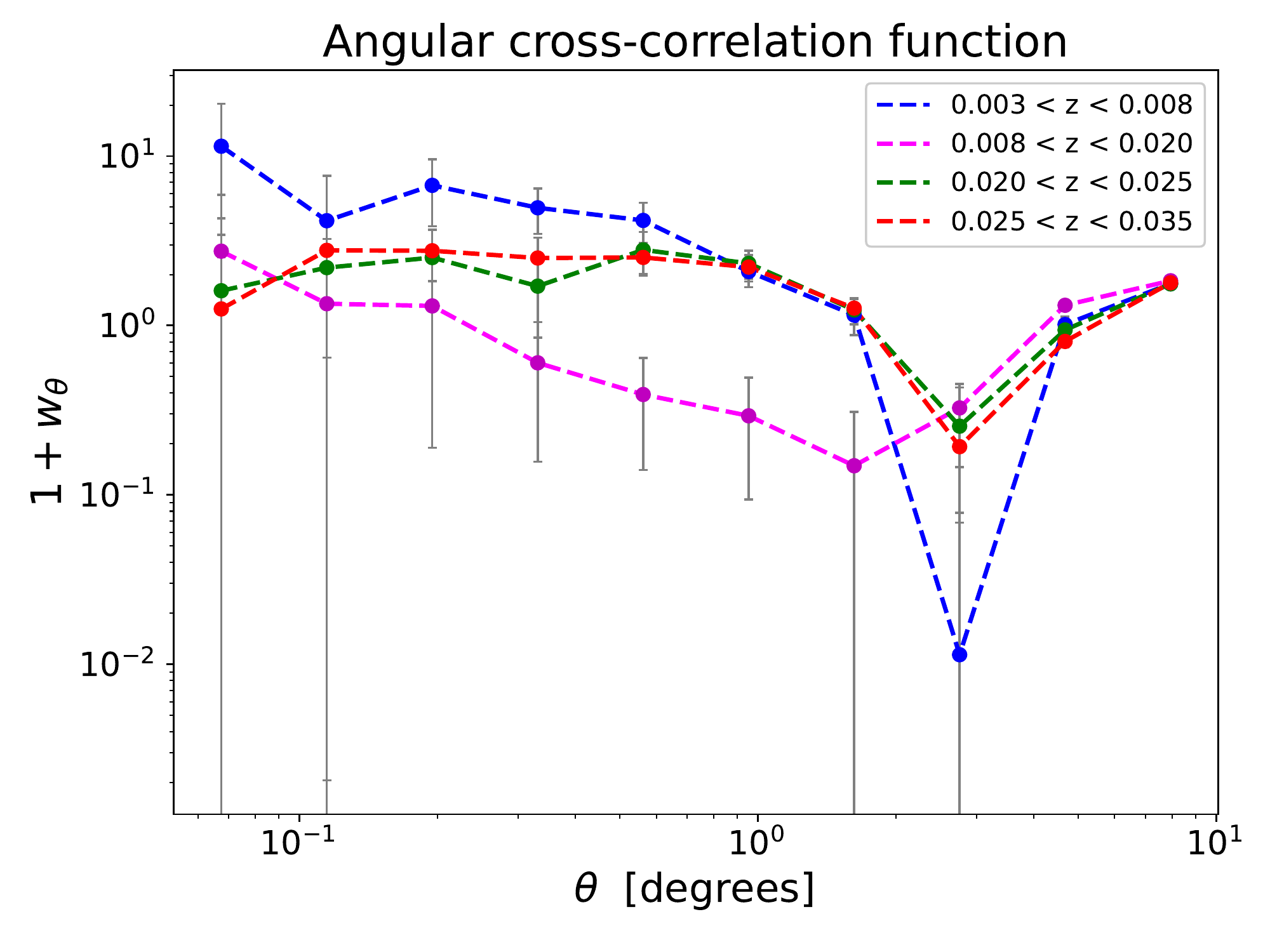}
        \includegraphics[width=1.\columnwidth]{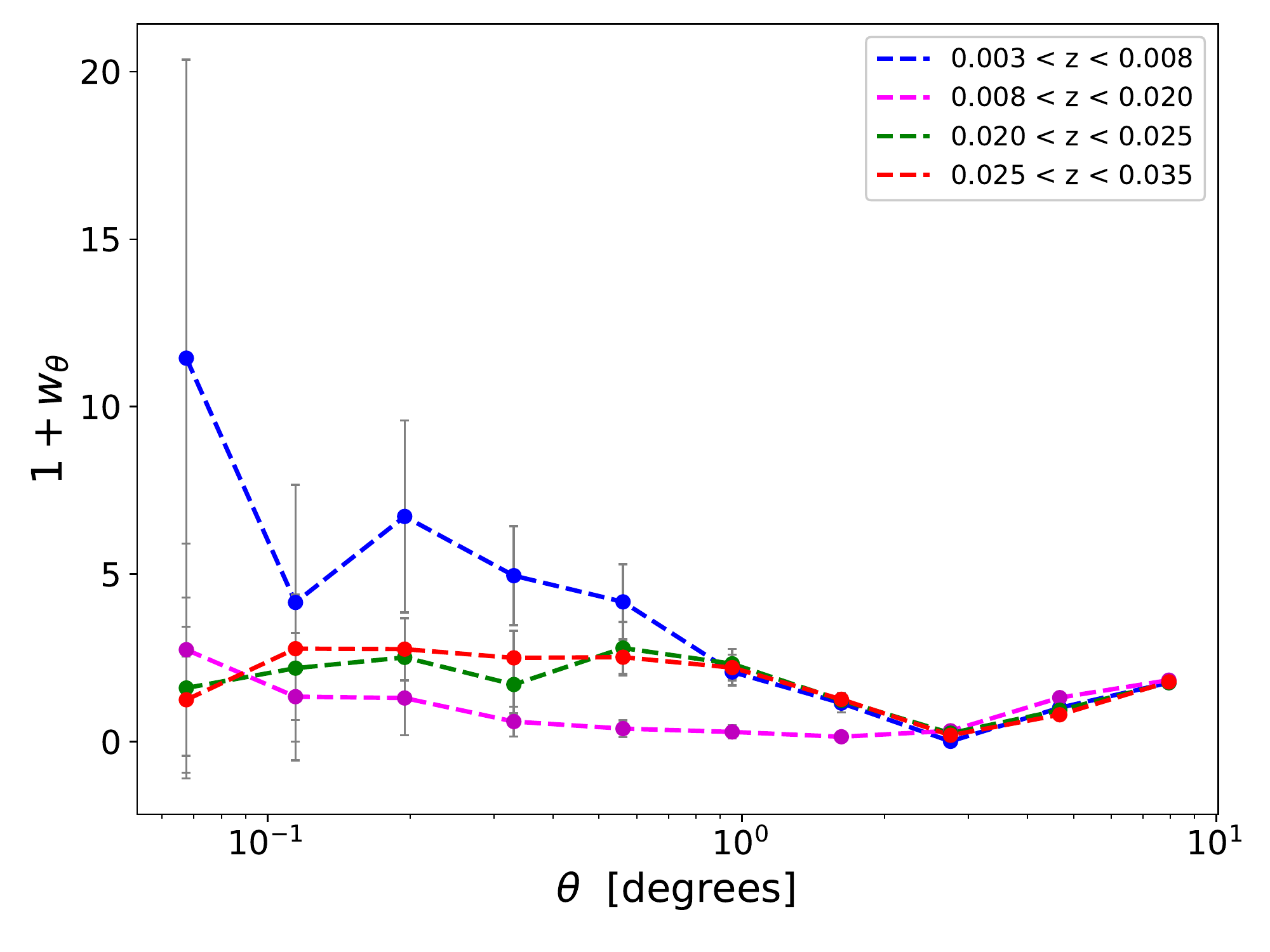}
    \caption{Two-point angular cross-correlation function for LSBGs and galaxies with different redshift intervals. {Left} panel is in log-log scale. {Right} panel shows the two-point angular CCF in semi-log scale for visualization purposes.}
    \label{fig:tpacc}
\end{figure*}

In order to confirm our results quantitatively we provide here a characterization of the spatial correlation of the structure through statistical parameters. We measure the spatial distribution of galaxies in two dimensions as projected onto the plane of the sky. This can be studied through the two-dimensional projected angular correlation function $\omega(\theta)$, which is defined through the expression

\begin{equation}
\label{2pac}
dP = N^2[1+\omega(\theta)]d\Omega_1 d\Omega_2
,\end{equation}

where N represents the surface density per squared radian and $d\Omega_1$, $d\Omega_2$ are infinitesimal solid angles separated by an angle $\theta$. This equation represents the probability with respect to a Poissonian distribution of finding two galaxies with an angular separation $\theta$. The usual estimation for $\omega(\theta)$ is given by the ratio of the number of pairs of galaxies counted in the sample to that expected from a random distribution with the same mean density and sampling geometry. 

The statistical characterization of the spatial correlation calculated here comprises two steps. First, the two-point angular correlation function was studied to compare cluster properties and identify the group whose statistical parameters are most compatible with those from LSBGs. In the second step, the two-point angular cross-correlation function (CCF) was calculated. Due to the low number density, the first step gave inconclusive results and so we limited the study to the two-point angular CCF between LSBGs and each of the groups of galaxies.

Using the samples of LSBGs and galaxies with estimated redshift described previously, we create 10 000 random samples (no noticeable difference was found when using 100 000) and we calculate the two-point angular CCF of the  LSBG group following \citep{1999MNRAS.305..547C}. We used the correlation function implemented in the mock\_observables sub-package from Halotools v0.7 \citep{2017AJ....154..190H} as described as:

\begin{equation}
\label{tpaccf}
1 + w(\theta) \equiv \mathrm{DD}(\theta) / \mathrm{RR}(\theta),
\end{equation}

where $\mathrm{DD}(\theta)$ and $\mathrm{RR}(\theta)$ are the number of sample pairs and of random pairs with separations equal to $\theta,$ respectively.

In Eq. \ref{tpaccf} we used the Landy-Szalay correction \citep{1993ApJ...412...64L} to the pair counts which is more unbiased than the natural method and produces a nearly Poissonian variance. 
Errors are estimated through a self-made\footnote{see \url{github.com/javier-iaa/LSBGs_1052_paper/Two_Point_Angular_Cross-Correlation_Function.ipynb} for details on the calculation} algorithm performing bootstrap re-sampling of the two-point angular CCF. A collection of 50 randomized catalogs populating the same sky coverage as the data are generated by bootstrap. The error is then estimated through the standard deviation for the statistic of the resulting set.
The two-point angular CCF obtained in this way is calculated for a number of bins corresponding to angular distances in degrees. The size of the bin is determined in general by the sample size. In our case, we calculated the two-point angular CCF with several spatial bins spaced logarithmically from 0.05 to 10 deg. For hierarchical structures, a decreasing trend of $1+\omega(\theta)$ with radius (i.e., $\theta$) is expected. 



Figure \ref{fig:tpacc} shows the two-point angular CCF for LSBGs and galaxies corresponding to each of the redshift peaks identified. We note that for $0.05<\theta<1.0$ the angular cross-correlation is higher for the first peak than for the rest. This is the interval where the highest density of galaxies is found and so it is the most relevant range in which to calculate the statistics. For smaller angular distances, the density is too small (zero counts in some cases), and for larger angular distances the two-point angular CCF converges.

These results indicate a higher spatial correlation between LSBGs and galaxies with spectroscopic measurements located in the 0.003~<~z~<~0.008 interval. Taking into account this correlation and that the number of LSBGs is expected to decrease strongly with distance when applying a selection cut in r$_{eff}$~>~5~arcsec, these results suggest that the majority of  detected LSBGs are associated with the structure located at the redshift interval 0.003~<~z~<~0.008 to which NGC~1052 belongs. {However, we note the infeasibility of distance estimates in individual galaxies, and only a statistical study of the complete sample of LSBGs is possible through this analysis.}

\subsection{Spatial distribution of outlier LSBGs in effective radius}\label{sec:4.3}

In Section \ref{sec:Properties} we analyze the properties of the LSBG sample. We identify an anomalous effective radius distribution in which we find a number of LSBGs with high outlier values in relation to the main declining distribution. Here, we explore {this issue further}.

\begin{figure*}
\centering
        \includegraphics[width=1.0\textwidth]{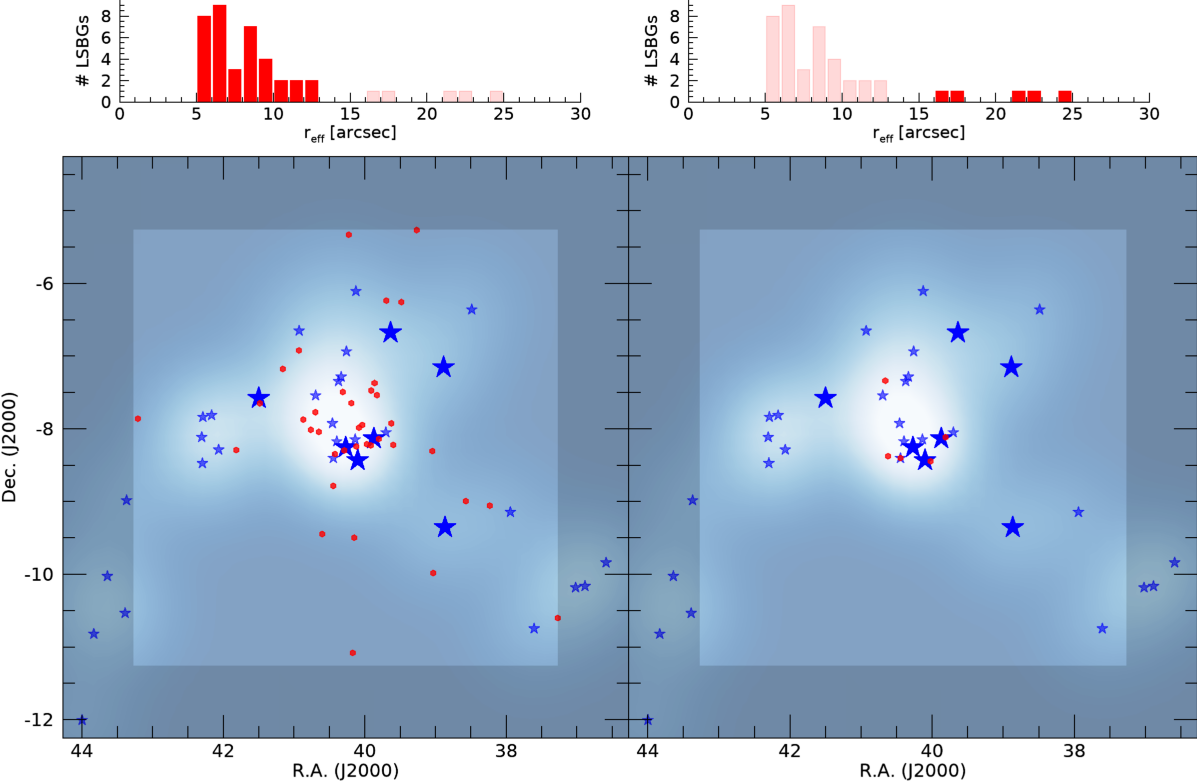}
    \caption{Spatial distribution of LSBGs and galaxies with spectroscopic measurements at 0.003 < z < 0.008, similar to that of Fig. \ref{fig:Spatial_distribution}. Here, LSBGs are plotted separately with r$_{eff}$ < 15 arcsec (left panel) and r$_{eff}$ > 15 arcsec (right panel). In the upper part of both panels, the distribution of effective radii of the LSBGs is presented with dark red for the galaxies plotted in their lower panel and light red for the hidden ones.}
    \label{fig:R_spa}
\end{figure*}

\begin{figure}
\centering
        \includegraphics[width=1.0\columnwidth]{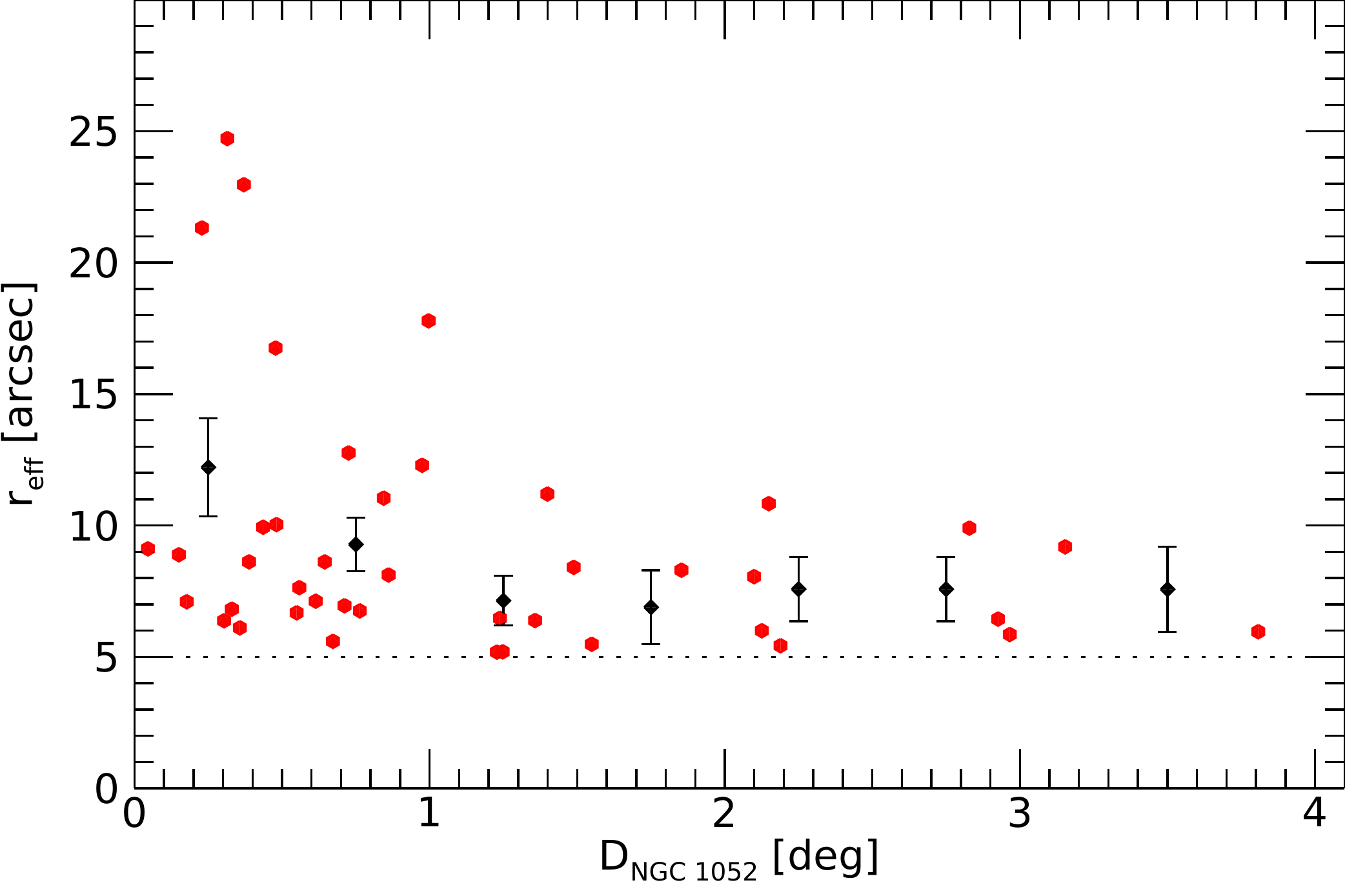}
    \caption{Correlation between the effective radius of the LBSGs and their projected distance to the NGC~1052 galaxy in red points. The black points with bars show the mean and the error in the mean in each bin of distance. The dotted line marks the detection limit of LSBGs of r$_{eff}$~=~5 arcsec.}
    \label{fig:R_vs_D}
\end{figure}

In Fig. \ref{fig:R_spa} we plot the LSBGs with an effective radius smaller (left panel) and larger (right panel) than r$_{eff}$ = 15 arcsec, together with the distribution of galaxies with spectroscopy located in the interval of 0.003 < z < 0.008, in the same way as performed in Fig. \ref{fig:Spatial_distribution}. As can be seen, the LSBGs with an effective radius larger than 15 arcsec are located in a compact area in the central region where the galaxy NGC~1052 is located. On the contrary, when plotting LSBGs with an effective radius of less than 15 arcsec, the LSBGs are homogeneously distributed over the entire region. Figure \ref{fig:R_vs_D} shows the correlation between the effective radius of the LSBGs and their distance in angular projection from NGC~1052 (D$_{NGC 1052}$). We find that LSBGs located near in projection to NGC~1052 have higher effective radius  on average: <r$_{eff}$>~=~12.2~$\pm$~1.9~arcsec for D$_{NGC 1052}$~=~[0,~0.5]~deg and <r$_{eff}$>~=~9.3~$\pm$~1.0~arcsec for D$_{NGC 1052}$~=~[0.5,~1.0]~deg, while we find <r$_{eff}$>~=~7.3~$\pm$~0.5~arcsec for D$_{NGC 1052}$~>~1 deg. Another way of looking at this is by exploring the values of D$_{NGC 1052}$ that LSBGs with higher effective radius have versus those with lower effective radius. We find that for LSBGs with r$_{eff}$ > 15 arcsec we obtain <D$_{NGC 1052}$>~=~0.48~$\pm$~0.14 deg against <D$_{NGC 1052}$>~=~1.25~$\pm$~0.15 deg for LSBGs with r$_{eff}$ < 15 arcsec.

These results show a statistically significant anomaly in which the LSBGs that have outlier values in the distribution of effective radii with r$_{eff}$ > 15 arcsec are clustered in the most central area explored, where NGC~1052 and the highest concentration of galaxies are located. {The difference between the effective radius of the LSBGs detected in D$_{NGC 1052}$ < 0.5 deg with respect to those detected in D$_{NGC 1052}$ > 1 deg is significant, almost a factor of two.}

\section{Discussion}

\begin{figure*}
\centering
        \includegraphics[width=1.0\textwidth]{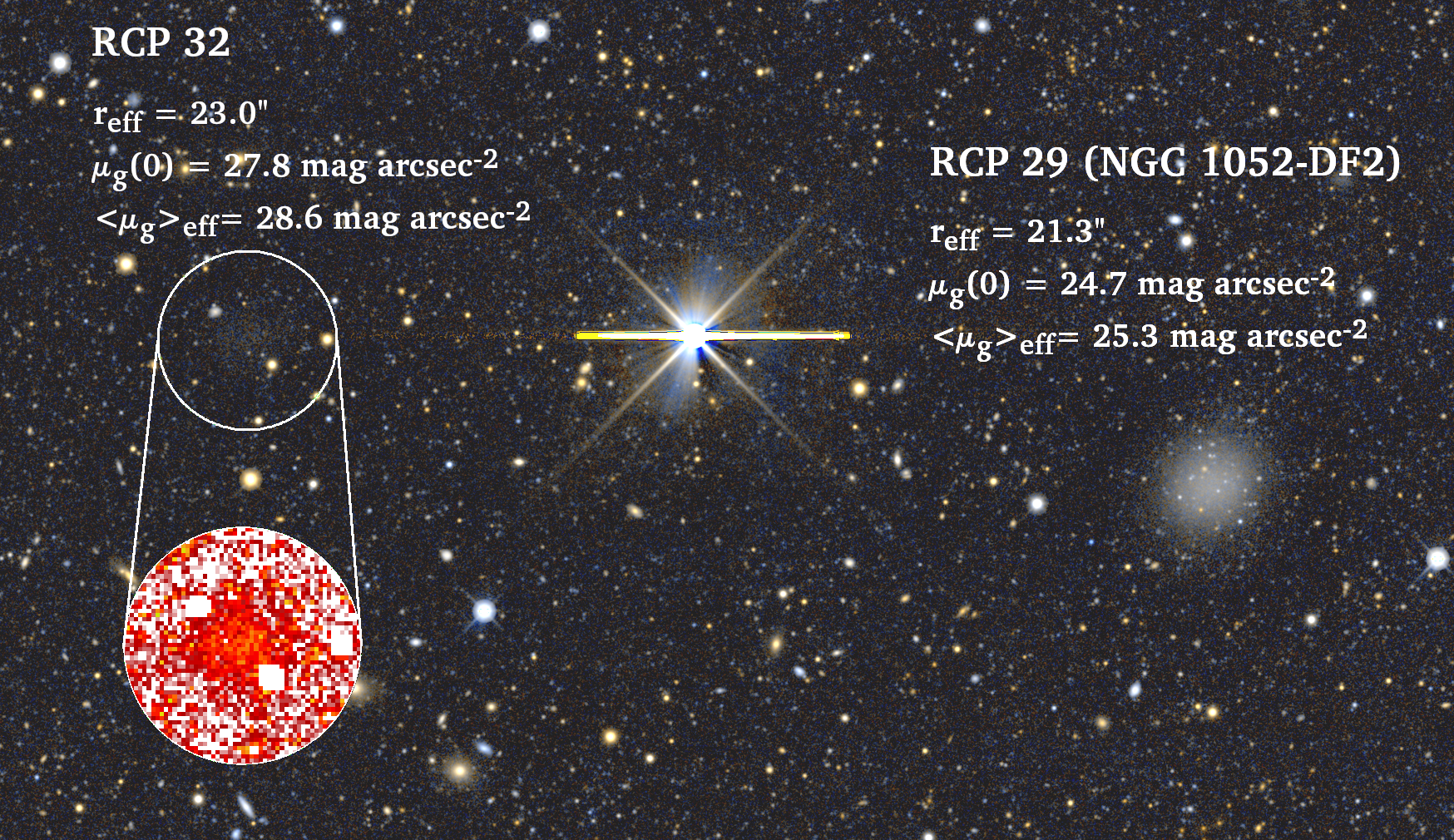}
    \caption{Color-composed image using the \textit{g} and \textit{r} bands of a field showing the proximity between RCP~32 and RCP~29 (NGC~1052-DF2). Given the low surface brightness of RCP~32, we zoom into a circular region of the processed image in which LSBGs were detected to improve the visualization of the galaxy (see Section \ref{sec:detection}).}
    \label{fig:DF2_cia}
\end{figure*}

In this study, we carried out an exhaustive and systematic detection of LSBGs in the environment of the galaxy NGC~1052. Photometric data from the Dark Energy Camera Legacy Survey {and} a dedicated pipeline for the detection of extremely faint objects allowed us to expand the catalog of LSBGs in this region to 42
objects,  20 of which are new objects. Among all the new objects, RCP~32 stands out with extreme properties: r$_{eff}$~=~23.0~arcsec and <$\mu_{g}$>$_{eff}$~$=$~28.6 mag arcsec$^{-2}$. This extremely low surface brightness makes RCP~32 one of the {lowest surface brightness} galaxies ever detected by integrated photometry, maybe only surpassed by the object BST1047+1156 located in the Leo I group \citep[][]{2018ApJ...863L...7M}. The clear presence of RCP~32 in the processed images by our pipeline with relatively high S/N (see Fig. \ref{fig:Binning}), the presence in both \textit{g} and \textit{r} bands (the \textit{z} band is not deep enough), the marginal presence in other data sets\footnote{RCP~32 appears marginally detected in the deep images by { \cite{2019A&A...624L...6M} and \cite{2021arXiv210909778K}}.}, and {its} overdensity of GC candidates {all confirm that} RCP~32 is a real object and not an artifact or reflection from nearby stars in the images. The fact that RCP 32 has remained undetected so far is remarkable given that NGC~1052-DF2, located {at} just 10 arcmin in projection (see Fig. \ref{fig:DF2_cia}), is a galaxy that has been extensively observed and analyzed in recent years. This result highlights the importance of deep data and adequate photometric processing in revealing the structures with the lowest surface brightnesses. 

Our analysis of the available spectroscopy in this region has identified a structure at a redshift range of 0.003~<~z~<~0.008 with a size of approximately 1 Mpc in diameter assuming the distance of 20 Mpc at which NGC~1052 is located \citep[][]{2013AJ....146...86T}. The spatial distribution of the detected LSBGs correlates strongly with this structure, {while the spatial correlation with the structures identified in the background of the redshift space is lower}. LSBGs are usually low-mass galaxies, and so they tend to be satellites of more massive galaxies; a spatial correlation is therefore expected. However, isolated or field LSBGs are also often found \citep[e.g.,][]{2021MNRAS.500.2049P}. The expected correlation between density and morphology of low-mass galaxies in group environments \citep[e.g.,][]{2006MNRAS.366....2W, 2011ApJ...726...98K} might improve our study performed in Sect. \ref{sec:4.1} through bayesian analysis by including the morphology factor in the prior distribution.


However, before conducting a more sophisticated study on the spatial correlation, certain unknowns related with the redshift-independent distances in this region must be clarified. In particular, there is a debate about the distances at which the galaxies RCP 12 (NGC~1052-DF4) and RCP 29 (NGC~1052-DF2) are located, and accurate distances are crucial in determining dark matter content. If located at a distance of 20 Mpc, similar to that of NGC~1052, the stellar mass of these would be comparable to the dynamical mass \citep[][]{2018Natur.555..629V, 2019ApJ...874L...5V}. However, \cite{2019MNRAS.486.1192T}, \cite{2019ApJ...880L..11M} and \cite{2021MNRAS.504.1668Z} claim a significantly closer distance, which would alleviate the anomalies of these galaxies related to their dark matter content, and also their overluminous population of GCs \citep[][]{2018ApJ...856L..30V, 2021ApJ...909..179S}.   

{Certainly, our work provides observational clues that are interesting in this discussion. The analysis carried out in section \ref{sec:4.3} shows that the effective radii of the LSBGs in our sample have a clear dependence on the environment, in that LSBGs located close to NGC 1052 tend to be much larger. Previous works analyzing the correlation between effective radius and environment for LSBGs or dwarf galaxies show inconsistent results \citep[][]{2016MNRAS.461L..82J, 2017MNRAS.468..703R, 2019A&A...625A.143V, 2019MNRAS.485.1036M, 2021MNRAS.507.6045C, 2021arXiv211000015K}. Even focusing on only those studies that find such a correlation, this correlation is very weak, {with variations of only a small fraction of the average values of the effective radius across different environmental densities}. Here we find a strong {correlation}, with the mean values of effective radii almost doubling in regions close to NGC~1052 compared to those further away. Therefore, these strong variations of the effective radii for LSBGs in the NGC 1052 environment can be considered an anomaly.} 

The existence of a significant number of LSBGs with {larger apparent sizes in a region of the space is a} clear indication of a bimodality in the distances for the detected LSBGs. {Given that the galaxy NGC~1042 is located at a distance of around $\approx$ 13 Mpc \citep[][]{2007A&A...465...71T, 2019ApJ...880L..11M}, closer than the larger structure associated with NGC~1052 at 20 Mpc, this is expected}. {An} important point here is that among the LSBGs that form this anomaly, RCP 12 (NGC~1052-DF4) and RCP 29 (NGC~1052-DF2) are included, which would imply that they are likely components of this closer structure in the line of sight.

{Recent works by \cite{2020ApJ...904..114M} and \cite{2021arXiv210909778K} show evidence of tidal interactions in RCP 12 (NGC~1052-DF4) and RCP 29 (NGC~1052-DF2). It {is compelling} to explore whether this anomaly in the sizes of the LSBGs could be related to the tidal distortions caused by the central galaxy NGC~1052. Under this hypothesis, it would be interesting to explore what makes the environment of NGC 1052 special for creating such expansion of the sizes of the LSBGs by tidal interactions but not in any other similar group of galaxies. One explanation could be that offered by \cite{2021arXiv210909778K}, namely that the absence or deficit of dark matter in RCP 12 (NGC~1052-DF4) and RCP 29 (NGC~1052-DF2) makes them more susceptible to being tidally disturbed. However, this would lead to other problems. The first is that not only do RCP 12 (NGC~1052-DF4) and RCP 29 (NGC~1052-DF2) appear with larger sizes in the vicinity of NGC~1052, but there are five LSBGs with an outlier effective radius in this region. This would create the need to find a characteristic  common to all of them that could explain their susceptibility to being tidally disturbed, and therefore their apparent larger sizes. Additionally, the tidal interactions observed in RCP 12 (NGC~1052-DF4) and RCP 29 (NGC~1052-DF2)\footnote{We note here that \cite{2021ApJ...919...56M} interpret the elongation on the outskirts of NGC~1052-DF2 as a disk.} appear as small deformations on the outskirts of these galaxies, something that in any case would not explain such a dramatic increase in the effective radius. Also, this hypothesis would not produce an immediate explanation for the overluminosity of the GCs observed in RCP 12 (NGC~1052-DF4) and RCP 29 (NGC~1052-DF2), which is another anomaly that needs to be taken into account. Further studies or models of the effects from tidal interactions, assuming a deficiency in dark matter for the LSBGs, would be interesting, perhaps allowing an explanation of all these circumstances in a common scenario.
}

{Indeed, GCs are of utmost importance in this discussion}, and the potential GC system detected in RCP 32 could be of significant interest. It seems to have a peak in its GCLF similar in luminosity to those of RCP 12 (NGC~1052-DF4) and RCP 29 (NGC~1052-DF2), and therefore overluminous if located at 20 Mpc. We note that RCP~32 is also one of the LSBGs in our sample with r$_{eff}$~>~15 arcsec. Assuming stellar properties for RCP 32 similar to those of RCP 29 (NGC~1052-DF2) with age~=~8.9~Gyr and [M/H]~=~-1.1 \citep[][]{2019A&A...625A..77F, 2019MNRAS.486.5670R}, the stellar mass for RCP~32 would be M$\star$ = 1.5$\times10^{7}$ M$\odot$ at 20 Mpc or M$\star$ = 6.1$\times10^{6}$ M$\odot$ at 13 Mpc. This is of the order of one-tenth of the stellar mass of RCP 12 (NGC~1052-DF4) or RCP 29 (NGC~1052-DF2). Given its low baryon content and its diffuse morphology, it is difficult to understand that RCP~32 could exist without the presence of a significant amount of dark matter. The mass threshold for survival of dark-matter-poor dwarf galaxies is estimated to be around 10$^8$ M$\odot$ \citep[][]{2006A&A...456..481B}. As no HI is detected in RCP 32, only the presence of dark matter would allow self-gravitation and long-term survival, and so the presence of dark matter seems necessary. This scenario is of course speculative, but if the overluminosity of their GCs and the presence of dark matter are confirmed, {being located at 20 Mpc} this would be a counterexample for the hypothesis that in the cases of RCP 12 (NGC~1052-DF4) and RCP 29 (NGC~1052-DF2) the overluminosity of their GCs is related to their absence of dark matter \citep[e.g.,][]{2021arXiv210308610T, 2021MNRAS.506.4841T, 2021ApJ...917L..15L}.

Finally, direct distance measurements for RCP 12 (NGC~1052-DF4) and RCP 29 (NGC~1052-DF2) using the tip of the red giant branch method have also been debated, reaching different results even through the use of identical data ({see \cite{2018ApJ...868...96C, 2018ApJ...864L..18V} and compare with  \citet{2019MNRAS.486.1192T, 2019ApJ...880L..11M}}). It is worth noting here the latest results using data of exceptional depth from the Hubble Space Telescope by \cite{2020ApJ...895L...4D} for NGC~1052-DF4 and \cite{2021ApJ...914L..12S} for NGC~1052-DF2. These authors find a distance of approximately 20 Mpc, {  suggesting that these galaxies would have indeed a strong deficit of dark matter and an overluminous population of GCs}. One of the points made by these latter authors is that the blending of sources by crowding could have a significant impact on the calculation of the tip. It is interesting to note that in the case of RCP 32, given its extremely low surface brightness, the crowding effect would be very low. We estimate that the stellar density of RCP 32 would be more than ten times lower than the cases of NGC~1052-DF2 and NGC~1052-DF4, with an accompanying difference in central surface brightness of approximately 3 mag arcsec$^{-2}$ . This is interesting because it would mean that similar observations in RCP~32 would have a more reliable tip of the red giant branch distance estimate {than for NGC~1052-DF2 and NGC~1052-DF4}, at least a priori. All these arguments make RCP 32 an object of great interest and follow-up observations could provide useful insights regarding the exotic properties of the LSBGs in this region.

\begin{acknowledgements}
{We thank the anonymous referee for interesting suggestions that improved this work.} We also thank Johan Knapen, Ignacio Trujillo and Mireia Montes for useful comments. The authors acknowledge financial support from the grants AYA2015-65973-C3-1-R and RTI2018-096228-B-C31 (MINECO/FEDER, UE), as well as from the State Agency for Research of the Spanish MCIU through the “Center of Excellence Severo Ochoa” award to the Instituto de Astrofísica de Andalucía (SEV-2017-0709). JR acknowledges support from the State Research Agency (AEI-MCINN) of the Spanish Ministry of Science and Innovation under the grant "The structure and evolution of galaxies and their central regions" with reference PID2019-105602GB-I00/10.13039/501100011033. JPG acknowledges funding support from Spanish public funds for research from project PID2019-107061GB-C63 from the "Programas Estatales de Generaci\'on de Conocimiento y Fortalecimiento Cient\'ifico y Tecnol\'ogico del Sistema de I+D+i y de I+D+i Orientada a los Retos de la Sociedad". This project used data obtained with the Dark Energy Camera (DECam), which was constructed by the Dark Energy Survey (DES) collaboration. Funding for the DES Projects has been provided by the U.S. Department of Energy, the U.S. National Science Foundation, the Ministry of Science and Education of Spain, the Science and Technology Facilities Council of the United Kingdom, the Higher Education Funding Council for England, the National Center for Supercomputing Applications at the University of Illinois at Urbana-Champaign, the Kavli Institute of Cosmological Physics at the University of Chicago, Center for Cosmology and Astro-Particle Physics at the Ohio State University, the Mitchell Institute for Fundamental Physics and Astronomy at Texas A\&M University, Financiadora de Estudos e Projetos, Fundacao Carlos Chagas Filho de Amparo, Financiadora de Estudos e Projetos, Fundacao Carlos Chagas Filho de Amparo a Pesquisa do Estado do Rio de Janeiro, Conselho Nacional de Desenvolvimento Cientifico e Tecnologico and the Ministerio da Ciencia, Tecnologia e Inovacao, the Deutsche Forschungsgemeinschaft and the Collaborating Institutions in the Dark Energy Survey. The Collaborating Institutions are Argonne National Laboratory, the University of California at Santa Cruz, the University of Cambridge, Centro de Investigaciones Energeticas, Medioambientales y Tecnologicas-Madrid, the University of Chicago, University College London, the DES-Brazil Consortium, the University of Edinburgh, the Eidgenossische Technische Hochschule (ETH) Zurich, Fermi National Accelerator Laboratory, the University of Illinois at Urbana-Champaign, the Institut de Ciencies de l’Espai (IEEC/CSIC), the Institut de Fisica d’Altes Energies, Lawrence Berkeley National Laboratory, the Ludwig Maximilians Universitat Munchen and the associated Excellence Cluster Universe, the University of Michigan, NSF’s NOIRLab, the University of Nottingham, the Ohio State University, the University of Pennsylvania, the University of Portsmouth, SLAC National Accelerator Laboratory, Stanford University, the University of Sussex, and Texas A\&M University.
\end{acknowledgements}


\appendix

\section{Spectroscopically confirmed galaxies in 0.003 < z < 0.008}

\begin{table}
\centering
\caption{Spectroscopically confirmed galaxies in 0.003 < z < 0.008.}
{
\label{tab:specs}
\begin{tabular}{lccc}
\hline
Name   &  R.A. & Dec. & V rad \\
  & deg & deg & km s$^{-1}$ \\
\hline
        VV 525    &   36.5887   &  -9.84083   &   2109    \\
        MRK 1039   &   36.8865   &  -10.1656   &   2111    \\
        MRK 1042   &   37.0190   &  -10.1834   &   2133    \\
        DDO 023   &   37.6061   &  -10.7488   &   2102    \\
        SHOC 124   &   37.9417   &  -9.14658   &   1608    \\
        {}[KKS2000] 51   &   38.4877   &  -6.36005   &   1410    \\
        NGC 0988   &   38.8656   &  -9.35619   &   1510    \\
        NGC 0991   &   38.8862   &  -7.15443   &   1532    \\
        NGC 1022   &   39.6363   &  -6.67743   &   1453    \\
       LEDA 4014647   &   39.7021   &  -8.04937   &   1665    \\
       NGC 1035   &   39.8712   &  -8.13294   &   1241    \\
       NGC 1042   &   40.0999   &  -8.43354   &   1371    \\
       UGCA 038   &   40.1258   &  -6.10639   &   1325    \\
       NGC 1047   &   40.1368   &  -8.14767   &   1340    \\
       NGC 0961   &   40.2604   &  -6.93592   &   1295    \\
       NGC~1052   &   40.2700   &  -8.25576   &   1510    \\
       WISEA J024120.23-071705.9   &   40.3343   &  -7.28500   &   1663    \\
       LEDA 1020521   &   40.3724   &  -7.34611   &   1144    \\
       LEDA 1007217   &   40.3961   &  -8.17348   &   1530    \\
       NGC~1052-DF2   &   40.4450   &  -8.40333   &   1803    \\
       WISEA J024149.89-075530.4   &   40.4581   &  -7.92502   &   1372    \\
       WISEA J024246.88-073231.2   &   40.6953   &  -7.54210   &   1344    \\
       MCG -01-08-001   &   40.9283   &  -6.65139   &   1410    \\
       NGC 1084   &   41.4996   &  -7.57847   &   1407    \\
       {}[SB2012] 023   &   42.0664   &  -8.28792   &   1375    \\
       WISEA J024839.94-074848.2  &   42.1665   &  -7.81342   &   1465    \\
       NGC 1110   &   42.2899   &  -7.83754   &   1333    \\
       LEDA 1003250   &   42.2966   &  -8.47465   &   1430    \\
       WISEA J024913.38-080651.3   &   42.3059   &  -8.11505   &   1447    \\
       LEDA 4016901   &   43.3694   &  -8.98488   &   1425    \\
       LEDA 976989   &   43.3897   &  -10.5357   &   1507    \\
       NGC 1140   &   43.6399   &  -10.0278   &   1501    \\
       LEDA 3994671   &   43.8319   &  -10.8213   &   1572    \\
       IC 0271   &   43.9977   &  -12.0078   &   1607    \\
\hline
\end{tabular}
}
\end{table}

\end{document}